\documentclass[journal]{vgtc}                

\ifpdf
  \pdfoutput=1\relax                   
  \usepackage{graphicx}                
  \DeclareGraphicsExtensions{.pdf,.png,.jpg,.jpeg} 
\else
  \usepackage{graphicx}                
  \DeclareGraphicsExtensions{.eps}     
\fi%

\graphicspath{{figures/}{pictures/}{images/}{./}} 

\usepackage{microtype}                 
\PassOptionsToPackage{warn}{textcomp}  
\usepackage{textcomp}                  
\usepackage{mathptmx}                  
\usepackage{times}                     
\usepackage{cite}                      
\usepackage{tabu}                      
\usepackage{booktabs}                  
\usepackage{balance}
\usepackage{enumitem} 

\usepackage{soul} 
\usepackage{xcolor,colortbl} 
\usepackage{multirow}
\usepackage{array}
\def\revcolor{black} 
\newcommand{\rev}[1]{\textcolor{\revcolor}{#1}}

\onlineid{0}

\vgtccategory{Research}

\vgtcpapertype{please specify}

\title{Arigat\={o}: Effects of Adaptive Guidance on Engagement and Performance in Augmented Reality Learning Environments}

\author{Maheshya Weerasinghe\thanks{University of Primorska, Slovenia} \thanks{University of St Andrews, United Kingdom}, Aaron Quigley\thanks{University of New South Wales, Australia}, Klen Čopič Pucihar\footnotemark[1], Alice Toniolo\footnotemark[2], Angela Miguel\footnotemark[2], Matjaž  Kljun\footnotemark[1]}

\authorfooter{
\textit{Authors version.}

\textit{Contact authors: amw31@st-andrews.ac.uk and matjaz.kljun@upr.si.}
}
\shortauthortitle{Maheshya Weerasinghe \MakeLowercase{\textit{et al.}}: Paper Title}

\abstract{
Experiential learning (ExL) is the process of learning through experience or more specifically ``learning through reflection on doing''. In this paper, we propose a simulation of these experiences, in Augmented Reality (AR), addressing the problem of language learning. Such systems provide an excellent setting to support ``adaptive guidance'', in a digital form, within a real environment. Adaptive guidance allows the instructions and learning content to be customised for the individual learner, thus creating a unique learning experience. 
We developed an adaptive guidance AR system for language learning, 
we call Arigat\={o} (Augmented Reality Instructional Guidance \& Tailored Omniverse), which offers immediate assistance, resources specific to the learner's needs, manipulation of these resources, and relevant feedback. 
Considering guidance, we employ this prototype to investigate the effect of the amount of guidance (fixed vs.\ adaptive-amount) and the type of guidance (fixed vs.\ adaptive-associations) on the engagement and consequently the learning outcomes of language learning in an AR environment. The results for the amount of guidance show that compared to the adaptive-amount, the fixed-amount of guidance group scored better in the immediate and delayed (after 7 days) recall tests. However, this group also invested a significantly higher mental effort to complete the task. The results for the type of guidance show that the adaptive-associations group outperforms the fixed-associations group in the immediate, delayed (after 7 days) recall tests, and learning efficiency. The adaptive-associations group also showed significantly lower mental effort and spent less time to complete the task.
} 


\teaser{
  \centering
  \includegraphics[width=\linewidth]{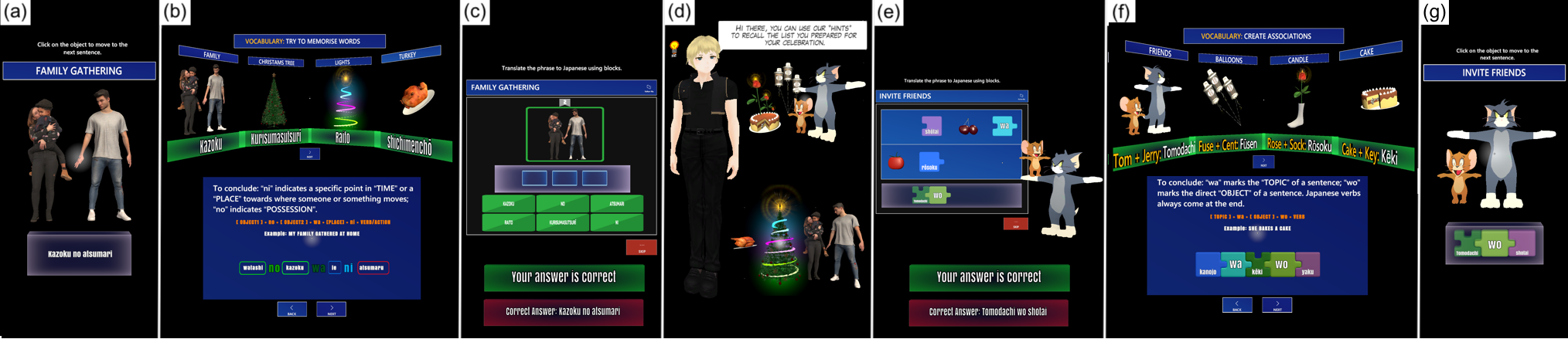}
  \caption{Arigat\={o} prototype and its experiential learning stages for two different topics: Christmas and birthday celebration. (a) Reflective observation (RO), (b) Abstract conceptualisation (AC), and (c) Active experimentation (AE), stages for fixed-associations condition on the Christmas celebration topic. (d) Concrete experience (CE) for both topics. (e) Active experimentation (AE), (f) Abstract conceptualisation (AC), and (g) Reflective observation (RO) stages for adaptive-associations condition on a birthday celebration topic. To avoid excessive clutter of the AR scene the black background was selected.}
  \label{fig:teaser}
}

\renewcommand{\manuscriptnotetxt}{}

\nocopyrightspace

\vgtcinsertpkg

\begin{document}


\firstsection{Introduction}
\label{sec:introduction}
\maketitle
\par Experiential learning (ExL) is a well-known learning approach used in education, training, facilitation, coaching and organisational development~\cite{lewis1994,kolb2001,ferrero2018}. ExL refers to the process of learning through experience or ``learning by doing''. 
One of the most influential models in ExL~\cite{lewis1994} is the Kolb's learning model~\cite{kolb1984}.
This model defines ExL as ``the process whereby knowledge is created through the transformation of experience. Knowledge results from the combination of grasping and transforming experience''~\cite{kolb1984}. The model is represented by four cyclical learning stages 
as seen in \autoref{fig:Kolb1984}: (1) concrete experience (feeling), 
(2) reflective observation (watching), 
(3) abstract conceptualisation (thinking), and 
(4) active experimentation (doing). 
One can enter the learning cycle at any stage, but all stages in the cycle must be addressed for  meaningful learning to occur. 

\par In learning, including experiential learning, guidance is beneficial. The term guidance has a broad meaning, but we adopt the generally understood definition of, guidance as ``any form of assistance offered to users so they can achieve a learning goal''. 
In more precise terms, the role of guidance is ``to simplify, provide a view on, elicit, supplant, or prescribe the scientific reasoning skills''~\cite{lazonder2016} involved in pursuing a goal. It can be explicit, as when a teacher provides instructions on how to solve a mathematical problem with all the in-depth explanations of concepts and skills students need to learn, or provides feedback on the current performance. Or it can be implicit, as it is provided within the environment for users to find, interpret and use in order to progress in the task at hand~\cite{anderson1985,canhoto2016}. In either case the amount and the type of guidance play an essential role in engagement and consequently, in the learning outcomes~\cite{carbonneau2015,Top2019,kalyuga2001l}. 

\par While several educational definitions of guidance exist~\cite{clark2009,lazonder2016,adams2008}, in the context of this study we focus on instructional ``guidance'', which is defined as ``providing users with accurate and complete procedural information (and related declarative knowledge) that they have not yet learned in a demonstration about how to perform the necessary sequence of actions and make the necessary decisions to accomplish a learning task''~\cite{clark2009}. Guidance also has different dimensions such as ``how much'' (amount) instructional support it provides~\cite{mayer2009,tobias2009,wise2009}, and ``what kind'' (type) of support it provides~\cite{adams2008}. Digital technologies such as augmented reality (AR) appear ideal to support adaptation of guidance. AR can provide an environment for simulating various experiences by combining the physical environments with in-context adaptive digital elements that can be manipulated by the user. \rev{AR can thus replicate the real-world ExL scenario (e.g. learning by decorating a room for Christmas celebrations with AR objects as can be done with real objects) and show information in a coherent and meaningful way within the real world context.}
Researchers have previously explored AR systems as tools to guide or teach different skills and activities that can be learned best through experience~\cite{huang2016,moorhouse2017}. Most of these projects have primarily focused on creating simulated environments to support different stages of the Kolb's experiential learning cycle~\cite{huang2016,moorhouse2017,moorhouse2019,vaughan2017,birt2018}. However, these projects explored generic guidance only, in which all the users receive the same set of general instructional cues, with the same amount, in the same way~\cite{birt2018,jarmon2009,mather2017,wei2016,lu2015}. Therefore, there are several questions that remain unanswered, such as: How to design and implement guidance in learning environments where the learning system needs to take over some or all of the tasks commonly conducted by educators? How the adaptation of guidance (the amount and the type) would affect the engagement and consequently the learning outcomes?

\par To this end, we built the Arigat\={o} \rev{-- an AR} prototype, which aims to effectively support learners to proceed through the ExL cycle facilitated by adaptive guidance. \rev{We deliberately selected AR as the most probable future technology that will be used in the classroom since it better supports real-world in-person communication and group collaboration compared to a desktop, tablet, and VR~\cite{rzayev2020effectsar,morrison2009beesar}. And rather than comparing the AR to other technologies our goal was to explore the design space of the AR.} With the Arigat\={o} prototype that provides options for a fixed and adaptive amount and type of guidance, we aimed to answer the following research questions: 

\begin{enumerate}[label=RQ\arabic*, leftmargin=22pt]
\item How do the dynamic adaptations of \rev{AR} guidance influence learners' (RQ1a) recall of previously learned information, (RQ1b) mental effort, (RQ1c) task completion time, and (RQ1d) instructional efficiency?
\item How does learners' engagement \rev{with AR content} in terms of (i.e., task completion time, mental effort and motivation) affect their performance (i.e., recall and efficiency)?
\end{enumerate}

\rev{The focus on AR in language learning covers the exploration of XR applications in the education and training domains making this study strongly relevant to the ISMAR community.} 


\section{Research Background}
\label{sec:relatedWork}
We first identify and analyse the relevant literature on experiential learning, instructional guidance and in particular the amount and type with a focus on language learning, user engagement, and language learning in AR. 

\subsection{Experiential Learning}
\par
\begin{figure}[hbt!]
    \centering
    \includegraphics[width=0.7\linewidth]{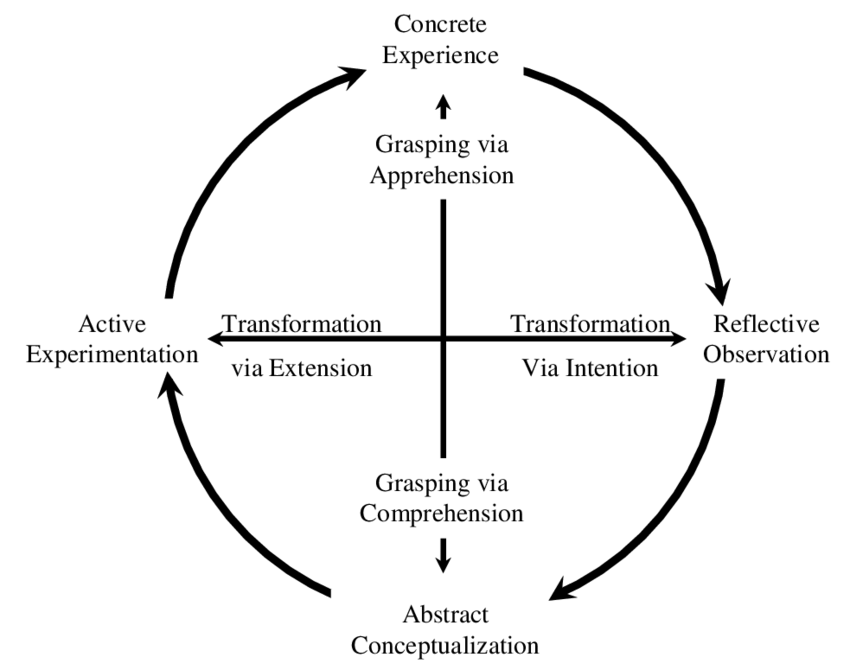}
    \caption{Experiential learning cycle by Kolb~\cite{kolb1984}}
    \label{fig:Kolb1984}
\end{figure}

 Experiential learning theory is based on the idea that learning is a process of acquiring knowledge through experience or ``learning  through  reflection  on  doing''~\cite{kolb1984,lewis1994}. Theoretical work on experiential learning continues to evolve and several models as ``theories-in-use'' have been proposed such as the Kolb's model~\cite{kolb1984}, Boud and Walker's model~\cite{boud1993}, Joplin's five stage model~\cite{joplin1981}, etc. Among these, the Kolb's experiential learning model is arguably the most influential model on educational scholarship~\cite{lewis1994}. 
As shown in \autoref{fig:Kolb1984}, the model describes experiential learning as a four-part process, where the learner is asked to engage in a new experience, actively reflect on that experience, conceptualise it, and integrate it with prior experiences and knowledge. After completing this cycle, future decisions can be based on the newly acquired concepts. 

\par A range of prior work has explored the use of simulated environments and AR technology to support experiential learning~\cite{huang2016,jantjies2018,moorhouse2017,moorhouse2019,vaughan2017,thoravi2019,mueller2003,birt2018,herbert2017,westerfield2015}. Most of these approaches have primarily focused only on the learning stages of the Kolb's experiential learning cycle, allowing learners to perform an action and providing feedback, which learners could reflect on~\cite{huang2016,moorhouse2017,moorhouse2019,vaughan2017,birt2018,grantcharov2004}. However, according to Kolb, the learner must continue cycling through all the four stages, thus creating a ``learning spiral of ever-increasing complexity'' in order to gain a better learning outcome~\cite{kolb1981}. Currently, we are unaware of any work that focuses on providing guidance to support learners move from one stage to another in the experiential learning cycle and progress on the learning spiral.

\subsection{Instructional Guidance}
\label{sec:instructional_guidance}
\par In the past, the instructional support provided during learning has been categorised and explained using different concepts and terms, such as learning paradigms (e.g.\ behaviourism, cognitivism and constructivism, etc.)~\cite{mergel1998,tobias2009}, instructional strategies (e.g.\ direct instructions, indirect instructions, interactive instructions, etc.)~\cite{merrill2002,weston1986} and instructional methods (e.g.\ lectures, case studies, peer feedback, quizzes, etc.)~\cite{cronbach1977,tobias1982}. Such paradigms, strategies and methods often emerge from different learning theories, and sometimes various belief systems and philosophies~\cite{cronbach1977,mergel1998,herman2009,wise2009,mayer2009}. Therefore, when trying to employ a construct, such as ``instructional guidance'', we depend on the fact that the use of different conceptual frameworks and theories tend to define and operationalise instructional support in different ways. In the following sections, we will focus on the amount and type of guidance as described in the literature.

\subsubsection{Amount of Guidance}
\label{sec:amount_of_Guidance}
\par Debates about the impact of instructional guidance and ``how much'' instructional support needs to be provided in a learning environment have been ongoing for at least the past century~\cite{mayer2009,kirschner2006,kuhn2007,tobias2009}. For example, is it better to instruct learners on what they need to know by presenting them the essential information, or is it better to allow learners to discover or construct essential knowledge for themselves? Koedinger and Aleven called this issue the ``assistance dilemma''~\cite{koedinger2007}, i.e., deciding whether to provide or withhold assistance. The contrast between the two practices can be better understood as a continuum. On one side of this continuum is the hypothesis that people learn best in an unguided or minimally guided environment~\cite{bruner1961,steffe1995}. On the other side is the hypothesis that learners should be given instructional guidance on the concepts and procedures required for a particular task~\cite{cronbach1977,mayer2009,kirschner2006}.

\par In a minimally guided learning environment, students are seen as active learners and are given opportunities to digest content for themselves rather than as passive learners who merely follow instructions. 
While these approaches have been adopted by some teachers and educators~\cite{wickens2015}, there have also been decades of efforts to discourage educators from using minimally guided learning approaches. The past half-century of empirical research on the topic has provided overwhelming and unambiguous evidence that minimal guidance during learning is significantly less effective and less efficient compared to guided learning~\cite{kirschner2006}. 
The superiority of the latter is explained in the context of human cognitive architecture and expert-novice differences~\cite{kirschner2006}. 
Wise and O'Neill present a review of evidence to support the view that the optimal amount of guidance is often somewhere in the middle of the aforementioned continuum and that the \textit{granularity} of the advice provided in a task (i.e.\ the level of details) is equally important~\cite{wise2009}. 

\subsubsection{Type of Instruction} 
\par Research on memory and learning has shown that comprehension and recall depend on different types of instructional methods and techniques that can be used to process and store information~\cite{dunlosky2013}. The mnemonic techniques have proven to be extremely effective in improving memory and recall, especially in the area of foreign language learning~\cite{raugh1975,paivio1981,cohen1980,amiryousefi2011}. Mnemonic is an instructional strategy designed for enhancing both memory and recall~\cite{worthen2011,putnam2015,mastropieri1991,pressley1982}. This technique connects new learning to prior knowledge through the use of visual and/or acoustic cues. The basic types of mnemonic techniques rely on the use of key words, rhyming words, or acronyms~\cite{putnam2015,mastropieri1991}.

\par In the field of language learning, mnemonics have mostly been used for vocabulary learning. Despite the great variety of techniques for presenting mnemonic, the ``keyword method''~\cite{atkinson1975} for creating mental associations to a known language has proven to be effective in the memorisation of vocabulary. In the keyword method, learners associate the sound of a word they want to learn to one they already know in either their first language or the target language. They then mentally create an image of the known word to memorise the association~\cite{pressley1982}. This association based technique provides a powerful tool for words that have a high degree of ``imagenability''~\cite{richardson1980}, or for word pairs between which the learner can form some kind of semantic link~\cite{ellis199}. The important aspect is that the keyword should clearly relate to the thing being remembered. This method also motivates learners to be more creative and use their minds more productively.

\subsection{User Engagement}
\par 
Engagement is a complex and multi-dimensional process~\cite{Sch2013,Pie2017,Top2019} intertwined with a learners’ internal indicators such as motivation, feelings, etc.\ (affective dimension)~\cite{fredricks2004}, mental effort, perceptions, etc.\ (cognitive dimension)~\cite{fredricks2004}, and observable actions such as performing various activities, interacting with a system, etc.\ (behavioural dimension)~\cite{App2008,fredricks2004}. In addition, Reeve and Tseng suggested incorporating agentic engagement as a fourth dimension of engagement~\cite{reeve2011}. Agentic engagement refers to the proactive and intentional activity of the learner to personalise the conditions of learning and to enrich external learning goals.

\par Previous studies have shown that behavioural, cognitive and affective dimensions predict learner’ performance both separately and in unison~\cite{alrashidi2016,Top2019}.  
This means that the dimensions of user engagement are interrelated and simultaneously affect human behaviour~\cite{fredricks2004}.

\par Kearsley and Shneiderman have stressed that engagement can be stimulated without technology, but digital technology opens up novel possibilities that are hard to achieve in a physical form~\cite{Kea1998}. Accordingly, research in this area has explored engagement with images~\cite{Fra2010}, video clips~\cite{Mur2008,Yaz2009}, music~\cite{Tak2003,Koe2012}, and real-life scenarios~\cite{Kat2008,Web2019}. 
However, while there are a limited numbers of studies such as~\cite{Top2019,guan2021}, there is a lack of research on learning and engagement in AR environments.

\subsection{Language Learning in AR}

\par A considerable body of literature focuses on guidance in AR environments to support learning in general~\cite{thoravi2019,birt2018,mather2017,wei2016,lu2015} and language learning in particular~\cite{ibrahim2018,dita2016,draxler2020,yang2018,arvanitis2021,seedhouse2014}. Prior work on language Learning in AR is summed in \autoref{tab:relatedwork}. We categorised it based on the following dimensions: (i) \textit{Hardware used:} mobile AR, HMD, sensors; \textit{Learning focus:} vocabulary and/or grammar; \textit{Guidance method:} generic or adaptive; and \textit{Learning method:} experiential, contextual, game-based learning, etc.~\cite{weerasinghe2019}). Importantly, most of the existing approaches deliver generic guidance, in which all users receive the same set of general instructional cues, and in the same way~\cite{birt2018,mather2017,wei2016,lu2015}. 

\begin{table}[htb!]
    \centering 
    \caption{Selected prior work related to language learning in AR environments and how our work differs alongside different dimensions.}
     \small
  \begin{tabular}{p{15mm} p{10mm} p{12mm} p{12mm} p{15mm}}
    \toprule 
    Study & \shortstack[l]{Hardware\\used} & \shortstack[l]{Learning\\Focus} & \shortstack[l]{Guidance\\Method} & \shortstack[l]{Learning\\Method} \\
    \midrule
    \shortstack[l]{Draxler et al.\\(2020)~\cite{draxler2020}} &
    Mobile & Grammar & Generic & Context-based learning  \\
    \shortstack[l]{Arvanitis et al.\\(2020)~\cite{arvanitis2021}} &
    Mobile & Vocabulary & Generic & Self-directed  learning \\
    \shortstack[l]{Yang \& Mei\\(2018)~\cite{yang2018}} &
    Mobile & Vocabulary & Generic & Game-based learning \\
   \shortstack[l]{Ibrahim et al.\\(2018)~\cite{ibrahim2018}} & 
    HMD & Vocabulary & Generic & Context-based learning \\
     \shortstack[l]{Vazquez et al.\\(2017)~\cite{vazquez2017}} &
    HMD & Vocabulary & Generic & Context-based learning \\
    \shortstack[l]{Dita\\(2016)~\cite{dita2016}} &
    Mobile & Vocabulary & Generic & Game-based learning \\
    \shortstack[l]{Seedhouse et al.\\(2014)~\cite{seedhouse2014}} &
    Sensors & Vocabulary & Generic & Experiential learning \\
    \shortstack[l]{Liu \& Tsai\\(2013)~\cite{liu2013}} & 
    Mobile & Vocabulary & Generic & Context-based learning\\
    \midrule
    \rowcolor{lightgray}
    \shortstack[l]{ Arigat\={o} (2022)} &
    HMD & Vocabulary Grammar & Adaptive & Experiential learning\\
    \bottomrule
    \end{tabular}
    \label{tab:relatedwork}
    \vspace{-0.3cm}
\end{table}

\par However, as pointed out in \autoref{sec:instructional_guidance}, the amount and the type of guidance play a vital role when providing instructional guidance. In addition, the adaptation of guidance to user needs is also important as noted in \autoref{sec:amount_of_Guidance}. Together they affect the engagement and consequently the learning outcomes of each individual learner. While there are studies such as~\cite{thoravi2019, huang2021} that provide adaptive guidance for performing a physical task, generally there is a lack of research on providing adaptive guidance based on the amount or type of instructions. Moreover, we are unaware of any work that focuses on the ``keyword method'' as the type of guidance used to support language learning in AR environments.


In order to address these gaps, we developed an adaptive guidance AR system for language learning called Arigat\={o}. We employ this prototype to investigate the effect of the amount (fixed vs.\ adaptive) and the type of associations (fixed vs.\ adaptive) on the engagement and consequently the learning outcomes of language learning in an AR environment. The research method to investigate the aforementioned effects is presented in the next section. 


\section{Research Method}
\label{sec:method}
\par 
The Arigat\={o} prototype for language learning was developed to answer our research questions. In this section, we present the prototype and we describe the study conditions, study design, study procedure, participants' sampling, data collection, and analysis.

\subsection{Study Conditions}
\label{sec:study_conditions}
\par The language selected for this study was Japanese, since it is unrelated to the Indo-European family of languages and we expected people would not be familiar with its grammar and vocabulary. We designed four different study conditions based on two different aspects of instructional guidance that can be adapted: the \textsc{amount} of guidance (\textsc{fixed-amount}, \rev{which is the same for all}, and \textsc{adaptive-amount}, \rev{which decreases based on the learners' performance)}, and the \textsc{type} of instructions, (\textsc{fixed-associations} \rev{using predefined 3D AR models of the vocabulary being learnt} and \textsc{adaptive-associations} \rev{using self selected 3D AR models to create associations (``keyword method'') to the vocabulary being learnt)}.

\par
\begin{figure}[t]
    \centering
    \includegraphics[width=1\columnwidth]{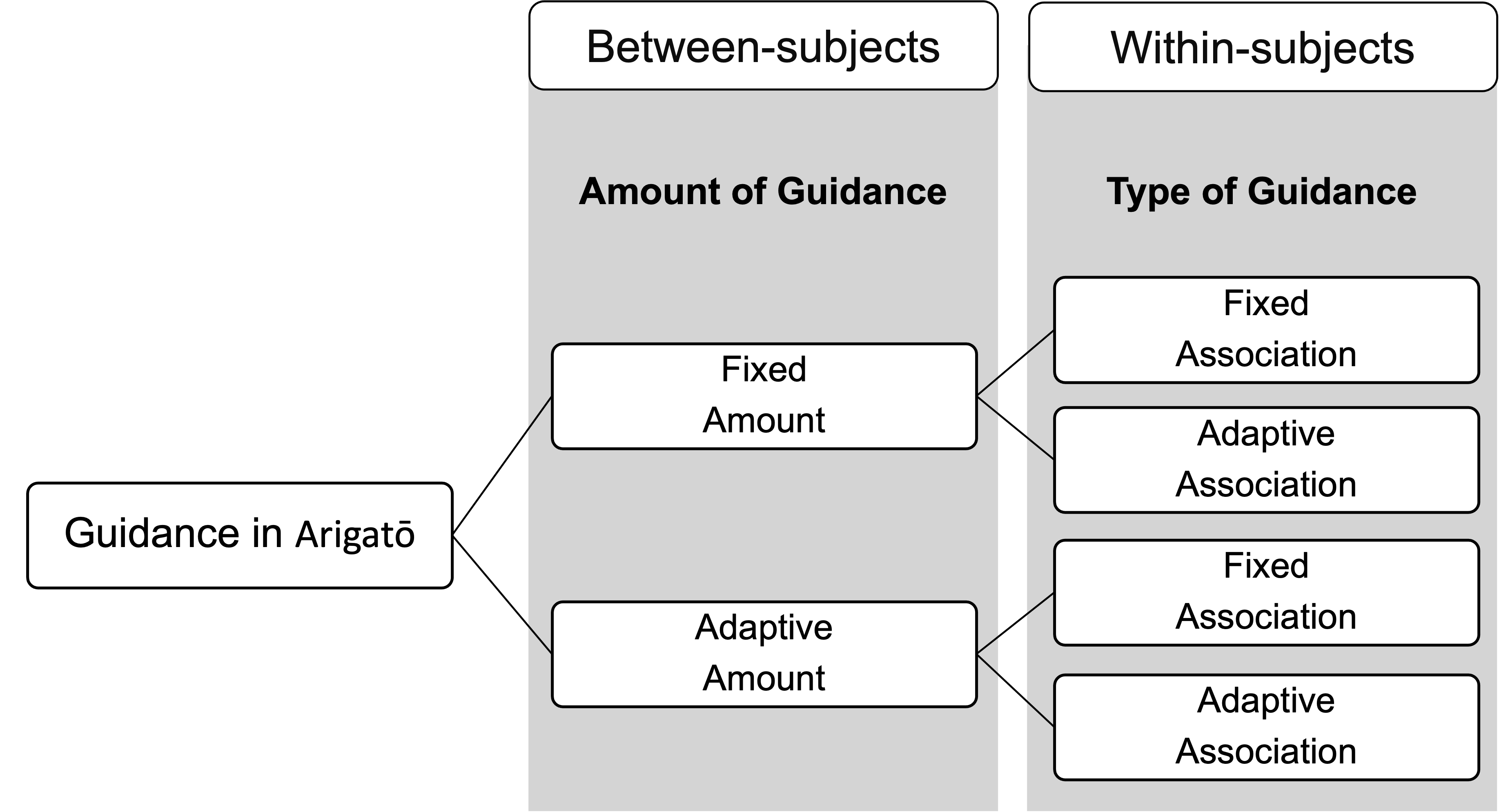}
    \caption{Study design and conditions.}
    \label{fig:design}
\end{figure}

 The structure of the design of the study with all four (4) study conditions is illustrated in \autoref{fig:design}.
The study was planned as a 2 x 2 mixed design study (including both within and between-subjects) envisaged to take approximately 60 to 75 minutes. \rev{A common within-subjects design would make this cognitively demanding learning study even longer (approx two to three hours), which might hinder participants’ performance and negatively affect the results. Other options such as splitting the study into several sessions would also introduce other biases (e.g. users might study between sessions, day to day performance might vary) and practical issues (e.g. getting all users back for the following session).}

The \textsc{amount} of guidance was thus evaluated as a within-subjects variable while the \textsc{type} of instructions as a between-subjects variable. This means that each participant either received the \textsc{fixed-amount} or the \textsc{adaptive-amount} of guidance, but all participants experienced both \textsc{fixed-associations} and \textsc{adaptive-associations}.

\par In each condition, participants had to learn Japanese vocabulary and grammar around a particular topic (Christmas or birthday celebration) by learning and understanding four (4) distinct phrases and their structure, and recall them successfully. The phrases related to Christmas were: family gathering, Christmas tree preparation, turkey dinner and lights decorations. The phrases related to birthday celebration were: invite friends, blow balloons, bake a cake and light up candles.

In the \textsc{fixed-amount} condition, 
\rev{at each stage of the learning cycle} participants received all the instruction for all the content needed to be learnt for all four (4) phrases repeatedly through consequent cycles until all phrases were correctly recalled. In the \textsc{adaptive-amount} condition, \rev{at each stage of the learning cycle} participants only received instructions for the phrases that were not recalled correctly in previous cycles. Thus, once a phrase was recalled \rev{correctly in the CE stage, the guidance for that phrase} was not shown in the next cycles. 
In the \textsc{fixed-associations} condition, participants were presented with \rev{predefined 3D AR models of objects corresponding to vocabulary of the phrases being learned}. In the \textsc{adaptive-associations} condition, the participants \rev{could self-select 3D AR models to create associations} for the corresponding vocabulary of the phrases being learned.

\par To avoid the ``order effects'' (the influence of the order in which the conditions are presented on participants' performances~\cite{schuman1981}), the order of \textsc{type} of instructions (\textsc{fixed} and \textsc{adaptive}) as well as the order of the topic being learnt (Christmas and birthday celebration) was balanced among the participants.

\subsection{The  Arigat\={o} Prototype}
\par The prototype was developed for the Magic Leap one AR head mounted display (HMD) with controller inputs~\footnote{https://www.magicleap.com/}, using the Unity3D game development environment~\footnote{https://unity.com/}. 
\rev{The Lumin SDK 0.26 and Magic leap tools were used for setting up the development environment with Lumin OS 0.98.3. The MRTK Mixed Reality tool kit~\footnote{https://docs.microsoft.com/en-us/windows/mixed-reality/mrtk-unity} was used for integrating the inputs and object manipulation techniques. Speech recognition was implemented with the IBM Watson speech to text API~\footnote{https://cloud.ibm.com/apidocs/speech-to-text}.}

The prototype replicates all four stages of Kolbs' experiential cycle. Before entering the cycle, the learner first sees an introduction window including all four (4) phrases related to the topic being learnt in English  in order to familiarise themselves with the intent and content of the study. Besides text instructions and explanations, audio instructions and explanations are also provided throughout the cycle.

\begin{figure}[hbt!]
    \centering
    \includegraphics[width=0.8\linewidth]{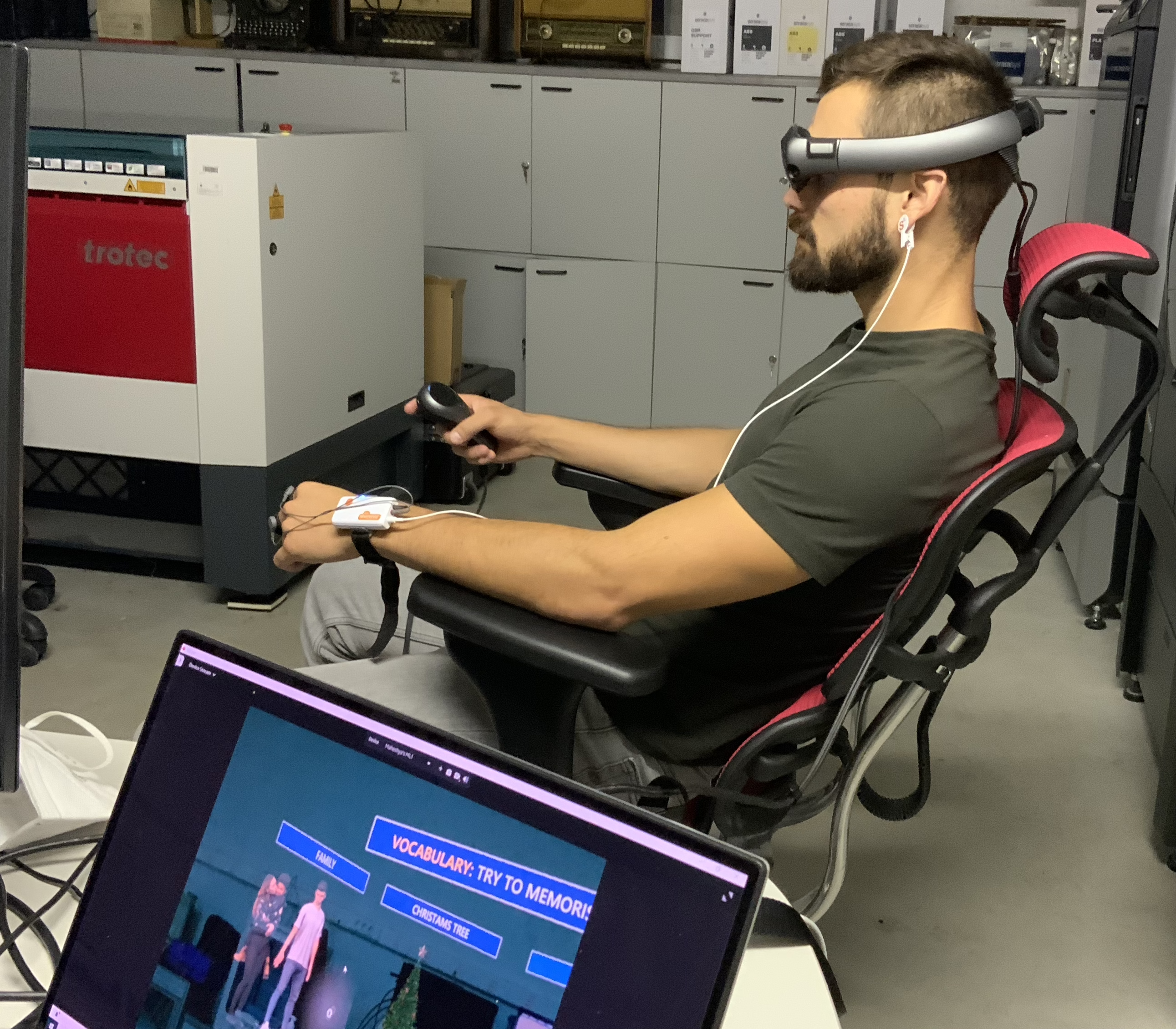}
    \caption{The study setup where the participant tries to finish the learning tasks visible in AR HMD and researcher follows the process on the laptop screen.} 
    \label{fig:scene01}
\end{figure}

\par The system starts the learning cycle in the abstract conceptualisation (AC) stage. The AC stage provides the description of the vocabulary and the basic grammar rules related to phrases being learnt. The vocabulary is presented by either (i) \textsc{fixed-associations} (for example, for the word lights -- ``raito'' in Japanese -- a 3D virtual object of a real world light set is shown by the word as visible in \autoref{fig:teaser}b), or (ii) \textsc{adaptive-associations} (for example, for the word candle -- ``rosoku'' in Japanese -- a merged 3D virtual object/a combined metaphor of a rose inside a sock is shown by the word as visible in \autoref{fig:teaser}f). After familiarising themselves with the content in the AC stage, the learner proceeds to the  active experimentation (AE) stage. 

\par In the AE stage, the system shows each English phrase with a puzzle task to generate the corresponding Japanese phrase made of three parts from six (6) possibilities offered. The puzzle is also presented in two ways based on the \textsc{type} of instructions: (i) a word puzzle with text components only (\autoref{fig:teaser}c) or (ii) a word puzzle together with 3D virtual objects for the learner to create associations (\autoref{fig:teaser}e) from a given set of objects. 
If the learner does not know the solution to the puzzle they can skip that phrase.
Feedback for the correct and incorrect phrases is \rev{given as text and highlights as shown in \autoref{fig:teaser}c and \autoref{fig:teaser}e. There, if the answer is incorrect, it highlights the given answer in red and shows the correct answer separately.}

\par After completing the puzzles related to all four phrases, the system nudges the learner to proceed to the concrete experience stage (CE). A 3D avatar is shown asking the learner to recall each phrase in Japanese  (\autoref{fig:teaser}d) by speaking it aloud. \rev{To evaluate the spoken answers, the IBM Watson speech recognition was not accurate enough, so a Wizard of Oz technique was used for the experiment. If the phrase is correctly voiced, the corresponding 3D AR model is displayed. The learner can manipulate the object and place it anywhere in the room as they would be able to do it in a real-world experiential learning scenario with physical objects.}

If all the phrases are not recalled correctly in the CE stage, the system automatically moves to the reflective observation (RO) stage. There, the learner \rev{can first go through all the phrases that include corresponding visual models and audio pronunciation, and then recall them by} selecting the correct Japanese phrase (out of four phrases) corresponding to the given English phrase (\autoref{fig:teaser}a and g). The system then returns back to the AC stage.  

In this way, the system nudges the learner to proceed further in the experiential learning cycle and consequently on the learning spiral until all four (4) phrases are recalled correctly in the CE stage. 
In the \textsc{fixed-amount} condition, all learners have to repeat all phrases in all stages in each cycle in the same way. In the \textsc{adaptive-amount} condition, the learner has to go only through and repeat previously incorrectly recalled phrases (in the CE stage) \rev{as the content related to previously correctly recalled phrases is no longer shown.}

One of the main features of the Arigat\={o} prototype is the possibility to move digital objects around the physical space as well as move various elements of the interface (e.g. puzzle) in order to complete the tasks given. It is this manipulation of the digital elements that can be brought in the visual field and removed that supports experiential learning or learning by doing.

\subsection{Participants}
\par In total, 28 participants were recruited for the study by invitation in a mailing list and web page as well as by snowball sampling. All were students and staff members from our university \rev{whose first language is one of the Indo-European languages. None of the participants had any prior knowledge in the Japanese language (identified via a short competency test questionnaire).} The between-subject experimental sample comprised 14 participants in the \textsc{fixed-amount} condition with 6 (46\%) females and 14 participants in the \textsc{adaptive-amount} condition with 8 (54\%) females. All participants were between 18 to 31 years old ($\overline{x} = 25$) and randomly assigned to one of the two conditions. The study was approved by the local Research Ethics Board.

\subsection{Procedure}

\par Participants were first given a consent form to sign, together with the Participant Information Sheet (PIS) outlining the entire research process, and were given an opportunity to ask any question related to the study. They were also instructed that they could abandon the study at any stage. Next, they were requested to fill out the Questionnaire on Current Motivation (QCM).

\par Before starting the actual task, they completed a five minute training session on a demo application to understand the interface and the interaction with the system. Participants were then instructed to complete a language learning task on one topic (Christmas or birthday celebration) in one condition (either \textsc{fixed-associations} or \textsc{adaptive-associations}) and the other topic in the other condition (see within-subject part of the \autoref{fig:design}). 
After finishing each topic, participants filled out a mental effort questionnaire and a recall questionnaire (to assess their immediate recall) asking participants to remember phrases they just tried learn. They were given a 5 minutes break in between the topics. 

\par In addition, at the end of the study, participants answered two standard questionnaires: a system usability (SUS)~\cite{lewis2009} and a user experience questionnaire (UEQ)~\cite{schrepp2017}. They also filled out a short post-questionnaire with demographic questions, questions about previous experience with AR technology, and questions about their vision. The whole experiment lasted from 60 to 75 minutes.

\par One week after the study, participants were again requested to answer the same recall questionnaire (as just after finishing the study) to assess their delayed recall. 

\subsection{Data Collection}
\par In all conditions, the performance progression data and the time stamp data were logged by the system as a part of behaviour engagement. To measure the motivation, the short form of the Questionnaire on Current Motivation (QCM) with 12 items/questions~\cite{rheinberg2001,freund2011} was used. QCM measures anxiety, challenge, interest, and probability of success on a five-point Likert scale ranging from 1 (``strongly disagree'') to 5 (``strongly agree''). Instead of focusing in individuals sub-dimensions (i.e., anxiety, challenge, interest, and probability of success), we used the mean score of the 12 items as an indicator of the overall motivation.

\par \rev{For measuring mental effort we used a standard 8-questions questionnaire from~\cite{joseph2013measuring} that focuses on mental load for learning contexts. To ensure comparability with other studies while maintaining coherence with our tasks, we just changed the wording to reflect the task at hand and better fit our learning scenario. In addition, we added 5 questions relevant to our study (How much effort did you invest in (i) memorising the words, (ii)  understanding particles, (iii) recalling the memorised words, (iv) building the sentences with puzzle blocks, and in (v) speaking the sentences out loud?). All 13 questions were answered on a nine-point scale. The mean scores of the 8 (MEQ8) and the 5 (MEQ5) answers as well as the overall mean score (MEQ13) were used as  indicators for the mental effort.}
 
\par The recall questionnaire (immediate and delayed) included Japanese phrases the participants practised during the learning tasks (e.g., ``How do you say `invite friends' in Japanese?'', ``How do you say `family gathering' in Japanese?'' with the result marked as either correct or incorrect).

\rev{To measure the usability of the system, we used the System Usability Scale (SUS) questionnaire~\cite{brooke1996} (10 questions answered on a five-point Likert scale). The SUS scores were calculated with the standard SUS analysis. 
 For measuring the user experience, we used the short version of the User Experience Questionnaire (UEQ-S)~\cite{schrepp2017,Ueq2019} with eight items/questions. The first four represent pragmatic qualities (Perspicuity, Efficiency and Dependability) and the last four hedonic qualities (Stimulation and Novelty)~\cite{Ueq2019}. The results are converted to a range between -3 to 3.}

\par To assess the reliability of motivation and mental effort questionnaires,  we performed Cronbach’s alpha test. Estimated reliability for each questionnaire (motivation Cronbach’s $\alpha = 0.71 $ and mental effort $\alpha = 0.89 $) was acceptable for the research purposes~\cite{ary2018}. 
For the recall questionnaire, the reliability was measured using the Kuder-Richardson 20 test~\cite{kuder1937} because of the binary nature of the results (correct/incorrect). The $KR = 0.76 > 0.5$ value indicates that the reliability of the recall questionnaire was also acceptable.

\subsection{Data Analysis}
\par 
Each data set collected in the study was first checked for normality using the Shapiro–Wilk normality test~\cite{sha1965}. 

\par The analysis was done in R studio, using ``WRS2'' R package. For immediate recall, delayed recall, mental effort, task completion time and learning efficiency, the statistical significance was examined using a mixed between-within subjects ANOVA on the 20\% trimmed means--``bwtrim''~\cite{mair2020}. The main between-subjects effect (group comparisons), the main within-subjects effect (e.g., due to repeated measurements), and the interaction effect, were computed using the ``sppba'', ``sppbb'', and ``sppbi'' functions respectively~\cite{mair2020}. Statistical significance for motivation was examined using the \rev{Mann-Whitney U test~\cite{sullivan2013}.} 

\par The resulting $p < 0.05$ are reported as statistically significant. All boxplots use a 1.5xIQR (interquartile range) rule and Tukey’s fences~\cite{int2016} for whiskers and identified outliers. Asterisk notation is used in figures to visualise statistical significance (ns: $p > 0.05$, *: $p < 0.05$, **: $p < 0.01$ and ***: $p < 0.001$). 

\par We also conducted a power analysis to check and validate the results and findings of the study. We calculated the effect size (Cohen's $d$) for each data set collected~\cite{cohen1988}, selected the minimum effect size (Cohen's $d = 1.251$) and estimated the statistical power ($1-\beta=0.8$) of data to check whether the type II error probability ($\beta$) is within an acceptable range for a given sample size ($n=14$ per group) and a significance level ($\alpha = 0.05$). The estimated power value $0.8$ shows that with the given sample size, we can have a 80\% chance that we correctly reject the null hypothesis with a significance level $0.05$.

\par Finally, we used a Pearson’s multiple correlation test~\cite{pla1983} to find out whether there were any correlations between a learner's mental effort, task completion time, motivation and their performances (i.e., recall and efficiency).



\section{Results}
\label{sec:result}
\par The results of the analysis based on the input from the 28 participants who completed the study is presented here. Before dwelling further into the effects of guidance on various measures we have checked whether the motivation has had an effect on the study. The results show that the motivation had no statistically significant effect on the rest of the results presented in this section \rev{(Mann-Witney U test, $U(N_{\textsc{fixed-amount}} = 14, N_{\textsc{adaptive-amount}} = 14) = 90.20$, $p > 0.05$)}. 

The following sections present the effect of adaptive guidance (amount and type) on learning performance (immediate and delayed recall), mental effort, task completion time, and learning efficiency (immediate and delayed). The last section presents correlations between these measures. 

\begin{figure*}[htb!]
    \centering
    \includegraphics[width=0.32\textwidth]{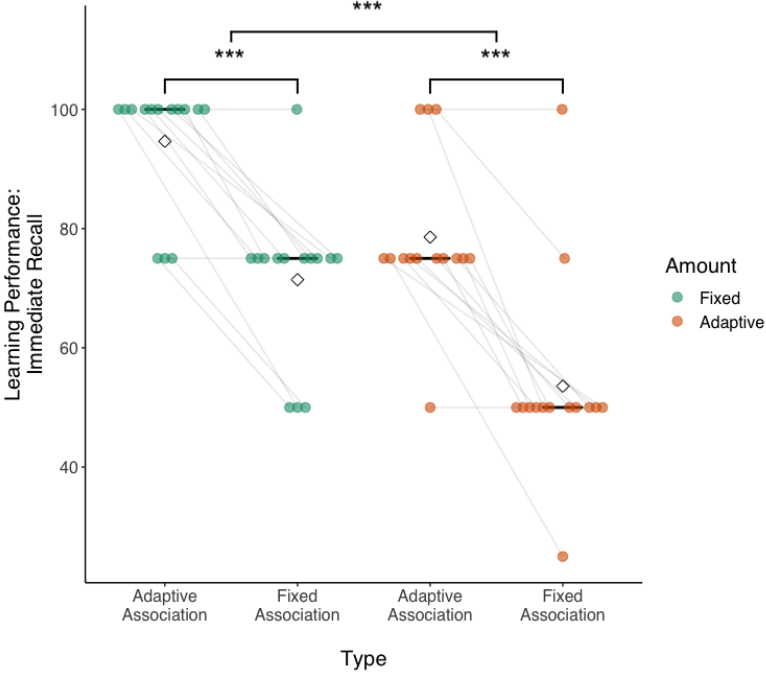}
    \includegraphics[width=0.32\textwidth]{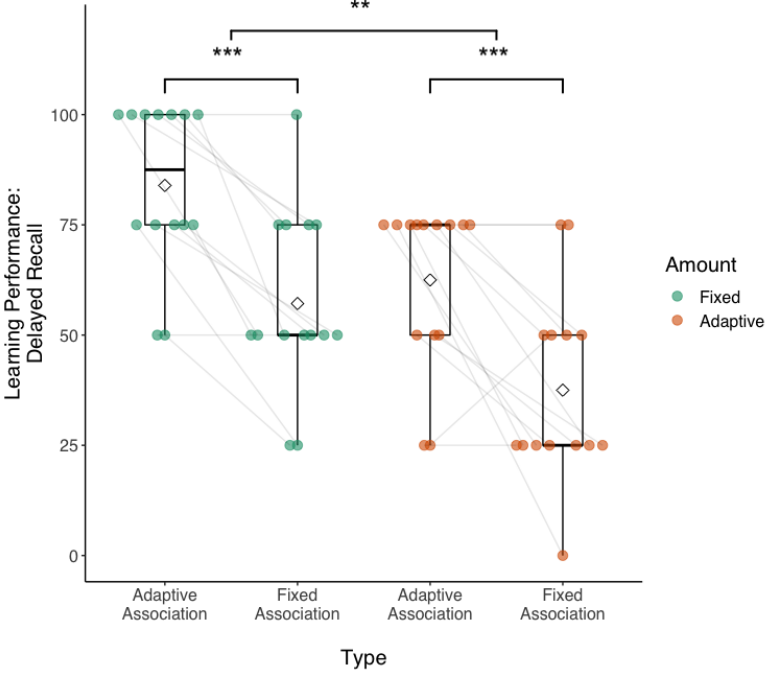}
    \includegraphics[width=0.32\textwidth]{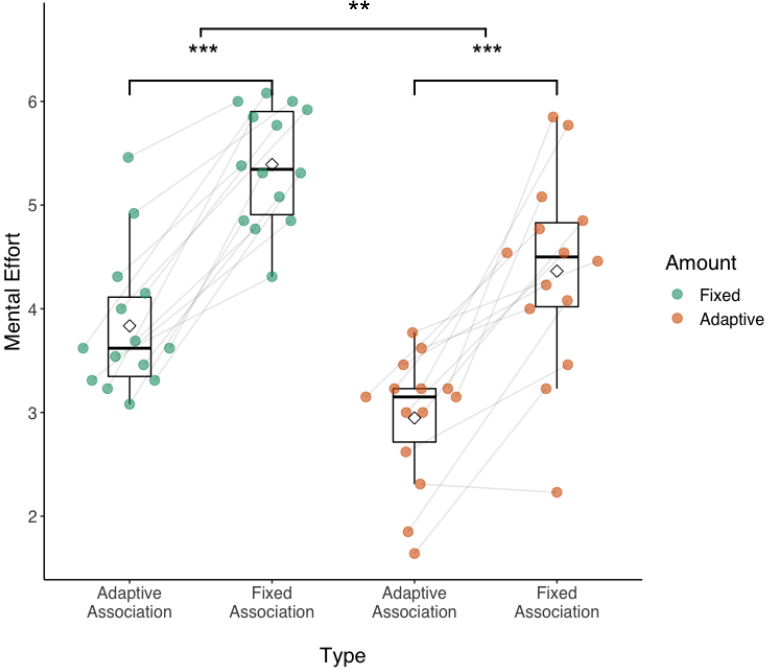}\\
    \includegraphics[width=0.29\textwidth]{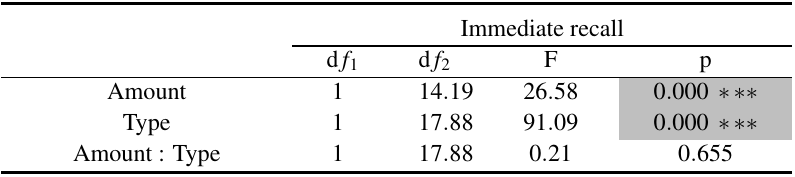}~~~~~
    \includegraphics[width=0.29\textwidth]{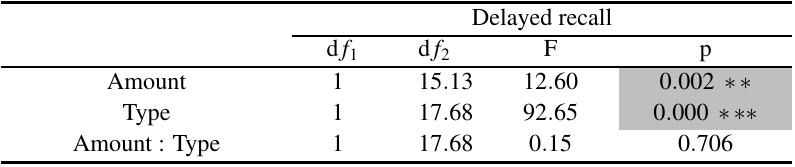}~~~~~~~~~
    \includegraphics[width=0.29\textwidth]{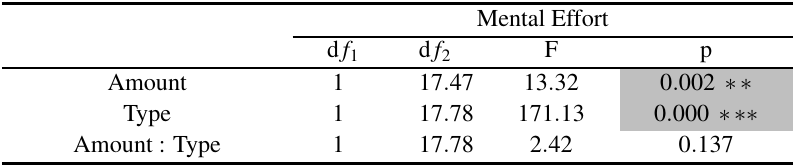}    
    \caption{
    Left: Immediate recall in percentage of correctly remembered phrases. 
    Centre: Delayed recall in percentage of correctly remembered phrases.
    Right: Overall mental effort (MEQ13) invested during the task.
    Below are results form mixed between-within subjects ANOVA on the 20\% trimmed means examining the between-subjects (\textsc{amount}), the within-subjects (\textsc{type}) and, the interaction (\textsc{amount:type}) effects, over dependent variables.}
    \label{fig:newtablegraphs1}
\end{figure*}

\subsection{The Effect of Guidance on Learning Performance}

Learning performance was measured with the recall questionnaires right after the study (immediate) and after 7 days (delayed). The results are presented in the following two sections.  

\subsubsection{Immediate recall}
\par The effect of the adaptive guidance  -- i.e., the \textsc{amount} (\textsc{fixed} and \textsc{adaptive}) and \textsc{type} (\textsc{fixed} and \textsc{adaptive}) -- on immediate recall is shown in \autoref{fig:newtablegraphs1} (top left).
The data summarised in \autoref{fig:newtablegraphs1} (bottom left) is analysed using a mixed between-within subjects ANOVA on the 20\% trimmed means~\cite{mair2020}.

\par The results in \autoref{fig:newtablegraphs1} (bottom left) show that the effects of the \textsc{amount} and \textsc{type} of guidance on immediate recall are statistically significant. The effect of the \textsc{amount} of guidance alone ($\mathit{df} = 14.19$, $p < 0.001$) is significant. \autoref{fig:newtablegraphs1} (top left) indicates that the immediate recall in the \textsc{fixed-amount} condition ($\overline{x} = 83.04\%$, $SD = 16.44$) is significantly better compared to the \textsc{adaptive-amount} condition ($\overline{x} = 66.7\%$, $SD = 19.15$). 

\par Similarly, the effect of the \textsc{type} on immediate recall ($\mathit{df} = 17.88$, $p < 0.001$) is also significant. Furthermore, \autoref{fig:newtablegraphs1} (top left) indicates that the immediate recall in the \textsc{adaptive-associations} condition ($\overline{x} = 86.61\%$, $SD = 11.57$) is significantly better compared to the \textsc{fixed-associations} condition ($\overline{x} = 62.5\%$, $SD = 14.42$). The interaction effect of the \textsc{amount} and \textsc{type} on the immediate recall ($\mathit{df} = 17.88$, $p > 0.05$) is not statistically significant.

\subsubsection{Delayed recall}
\par The effect of the adaptive guidance (\textsc{amount} and \textsc{type}) on delayed recall is shown in \autoref{fig:newtablegraphs1} (top centre). 
The data summarised in \autoref{fig:newtablegraphs1} (bottom centre) is analysed using a mixed between-within subjects ANOVA on the 20\% trimmed means~\cite{mair2020}. 

\par The results show that the effects of the \textsc{amount} and \textsc{type} of guidance on delayed recall are statistically significant. The effect of the \textsc{amount} of guidance alone ($\mathit{df} = 15.13$, $p < 0.01$) is also significant. \autoref{fig:newtablegraphs1} (top centre) indicates that the delayed recall in the \textsc{fixed-amount} condition ($\overline{x} = 70.54\%$, $SD = 23.20$) is significantly better compared to \textsc{adaptive-amount} condition ($\overline{x} = 50.00\%$, $SD = 23.15$). 

\par The effect of the \textsc{type} on the delayed recall ($\mathit{df} = 17.68$, $p < 0.001$) is also significant. \autoref{fig:newtablegraphs1} (top centre) indicates that the delayed recall in the \textsc{adaptive-association} condition ($\overline{x} = 73.21\%$, $SD = 18.12$) is significantly better compared to \textsc{fixed-association} condition ($\overline{x} = 47.32\%$, $SD = 20.24$). The interaction effect of the \textsc{amount} and \textsc{type} on the delayed recall ($\mathit{df} = 17.68$, $p > 0.05$) is not significant.

\subsection{The Effect of Guidance on Mental Effort}
\par The effect of the adaptive guidance (\textsc{amount} and \textsc{type}) on mental effort is shown in \autoref{fig:newtablegraphs1} (top right). 
The data summarised in \autoref{fig:newtablegraphs1} (bottom right) is analysed using a mixed between-within subjects ANOVA on the 20\% trimmed means~\cite{mair2020}.

\par The results show that the effects of the \textsc{amount} and \textsc{type} of guidance on mental effort are statistically significant. The effect of the \textsc{amount} of guidance alone \rev{(MEQ13 $\mathit{df} = 17.47$, $p < 0.01$; MEQ08 $\mathit{df} = 16.09$, $p < 0.01$; MEQ05 $\mathit{df} = 14.73$, $p < 0.01$)} is significant. \autoref{fig:newtablegraphs1} (top right) indicates that the overall mental effort in the \textsc{adaptive-amount} condition ($\overline{x} = 3.66\%$, $SD = 1.06$) is significantly lower compared to the \textsc{fixed-amount} condition ($\overline{x} = 4.61$, $SD = 0.98$). 

The effect of the \textsc{type} alone is also significant \rev{(MEQ13 $\mathit{df} = 17.78$, $p < 0.001$; MEQ08 $\mathit{df} = 12.58$, $p < 0.01$; MEQ05 $\mathit{df} = 17.73$, $p < 0.01$)}. \autoref{fig:newtablegraphs1} (top right) indicates that the overall mental effort in the \textsc{adaptive-associations} condition ($\overline{x} = 3.39$, $SD = 0.63$) is significantly lower compared to the \textsc{fixed-associations} condition ($\overline{x} = 4.88$, $SD = 0.73$). The interaction effect of the \textsc{amount} and \textsc{type} on mental effort \rev{(MEQ13 $\mathit{df} = 17.78$, $p > 0.05$; MEQ08 $\mathit{df} = 12.58$, $p > 0.05$; MEQ05 $\mathit{df} = 17.73$, $p > 0.05$)} is not statistically significant.

\subsection{The Effect of Guidance on Task Completion Time}
\begin{figure*}[htb!]
    \centering
    \includegraphics[width=0.32\textwidth]{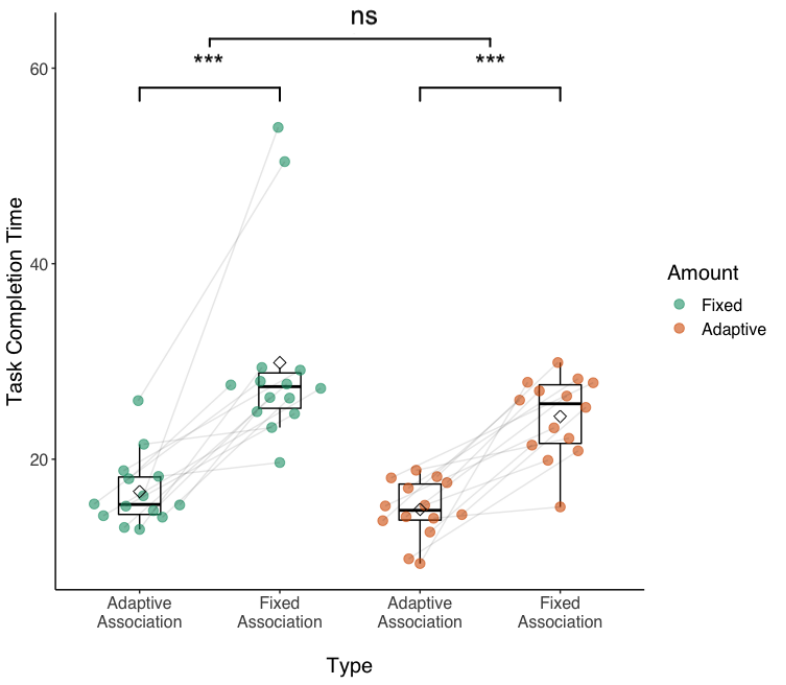}
    \includegraphics[width=0.32\textwidth]{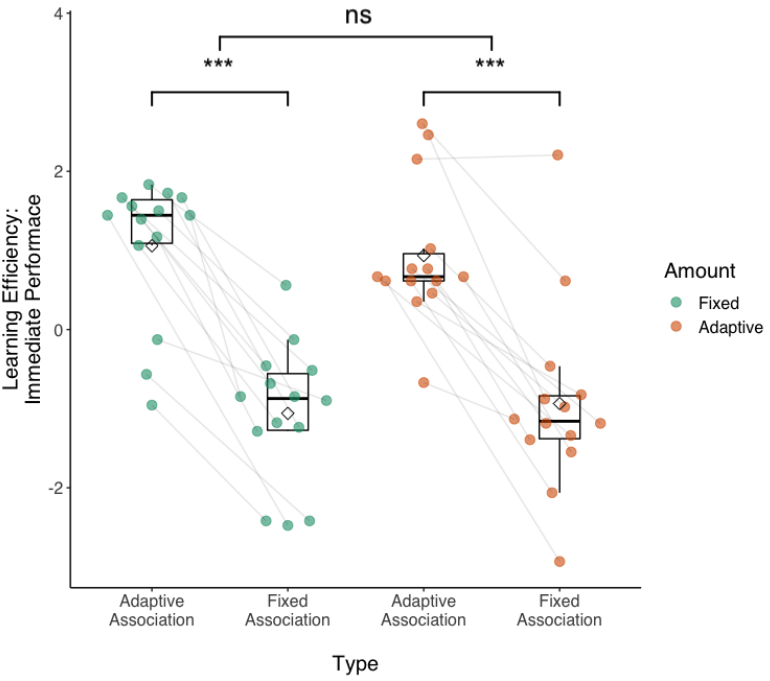}
    \includegraphics[width=0.32\textwidth]{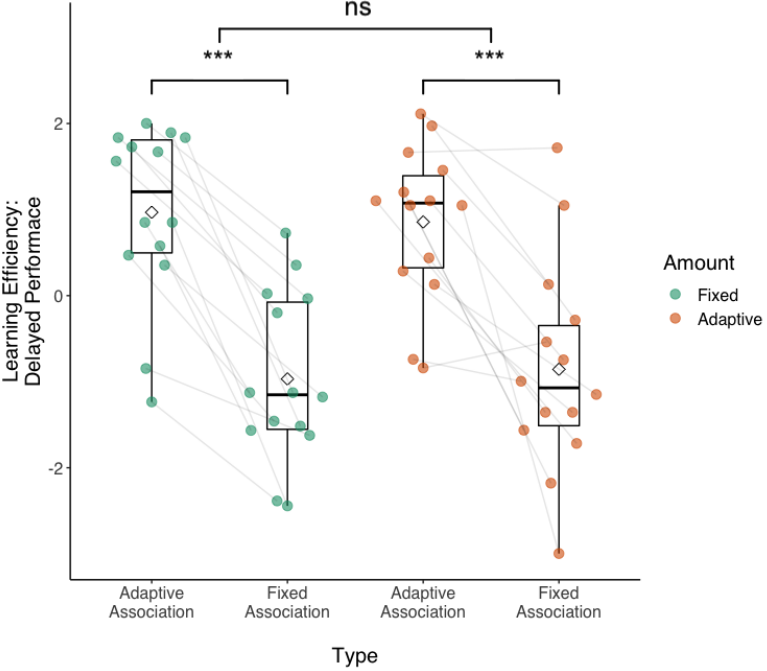}\\
    \includegraphics[width=0.29\textwidth]{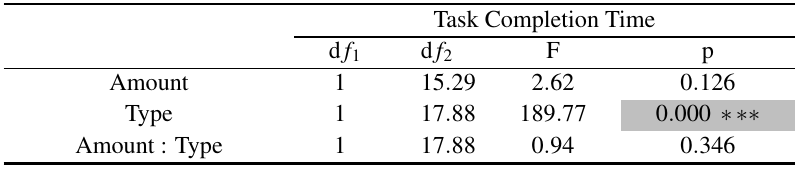}~~~~~
    \includegraphics[width=0.29\textwidth]{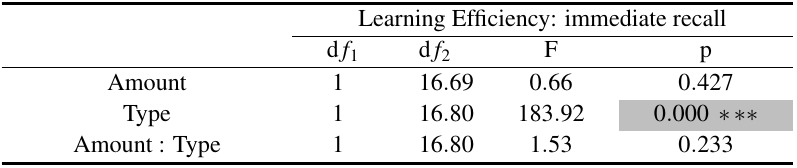}~~~~~~~~~
    \includegraphics[width=0.29\textwidth]{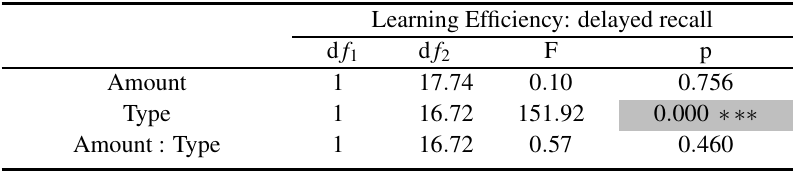}    
    \caption{
    Left: Task completion time in minutes. 
    Centre: Average learning efficiency for immediate performance.
    Right: Average learning efficiency for delayed performance.
    Below are results form mixed between-within subjects ANOVA on the 20\% trimmed means examining the between-subjects (\textsc{amount}), the within-subjects (\textsc{type}) and, the interaction (\textsc{amount:type}) effects, over dependent variables.}
    \label{fig:newtablegraphs2}
\end{figure*}

The results for the task completion time presented in \autoref{fig:newtablegraphs2} (bottom left) are analysed using a mixed between-within subjects ANOVA on the 20\% trimmed means. The results indicate that the \textsc{amount} of guidance has no statistically significant effect on the task completion time ($\mathit{df} = 15.29$, $p > 0.05$). It can thus not be concluded that participants in the \textsc{adaptive-amount} condition completed the task faster compared to participants in the \textsc{fixed-amount} condition. 

In contrast, the effect of the \textsc{type} on the task completion time is significant ($\mathit{df} = 17.88$, $p < 0.001$) as can be also observed in \autoref{fig:newtablegraphs2} (top left). This also shows that the task completion time in the \textsc{adaptive-associations} condition ($\overline{x} = 15.78~min$, $SD = 3.17$) is significantly lower compared to the \textsc{fixed-association} condition ($\overline{x} = 27.13~min$, $SD = 6.69$). The interaction effect of the \textsc{amount} and \textsc{type} on the task completion time is not statistically significant ($\mathit{df} = 17.88$, $p > 0.05$).

\begin{figure*}[htb!]
    \centering
    \includegraphics[width=0.47\textwidth]{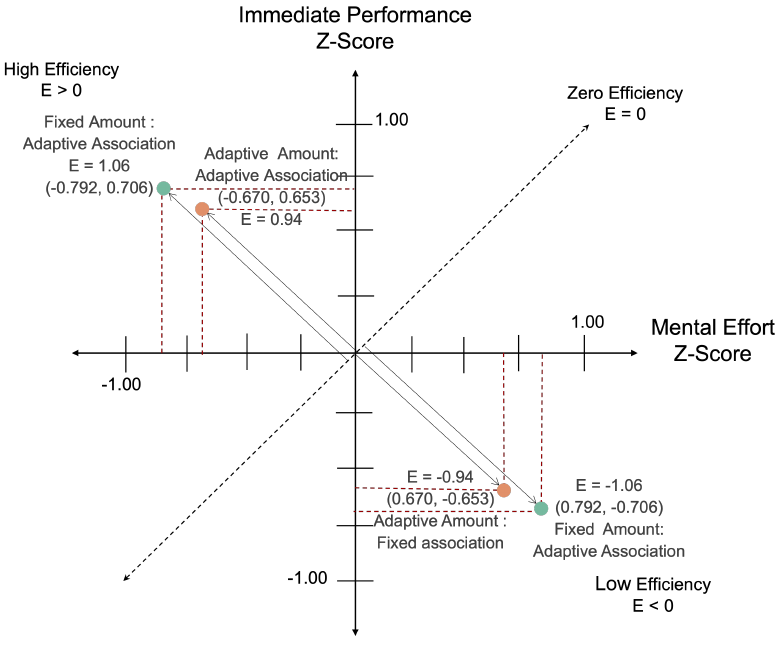}
    \includegraphics[width=0.47\textwidth]{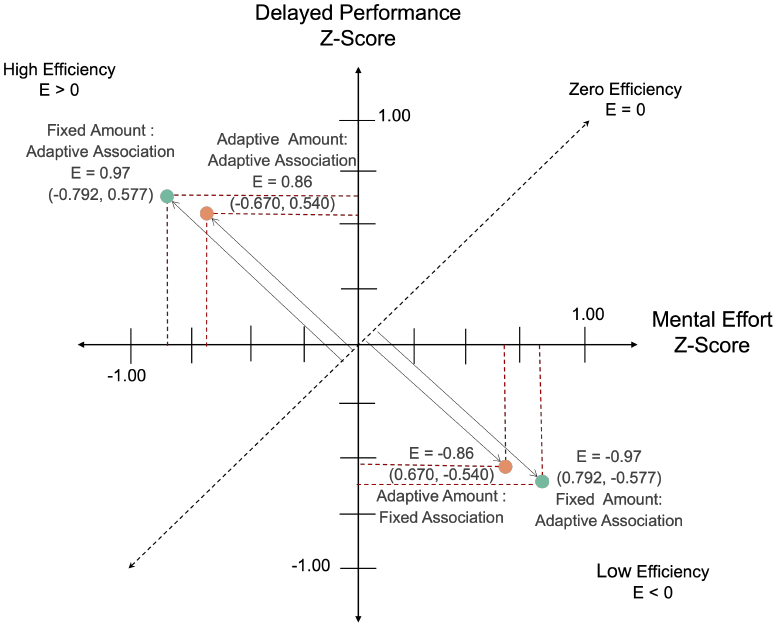}
    \caption{
    Left: Immediate recall learning efficiency. 
    Right: Delayed recall learning efficiency.
    }
    \label{fig:newtablegraphs3}
\end{figure*}

\subsection{The Effect of Guidance on Learning Efficiency}
\par The learning efficiency for each study condition was determined using \autoref{eq:2D}~\cite{paas1993,halabi2006,clark2011}. 

\begin{equation} \label{eq:2D}
E = \frac{z_P-z_M}{\sqrt{2}}
\end{equation}
\\
$E$ = Learning efficiency \\
$z_P$ = Average performance in Z-scores\\
$z_M$ = Average task difficulty in Z-scores\\

\par We measured the performance of each study condition based on the recall scores participants obtained after completing the task for that study condition. We estimated the difficulty of the task based on the mental effort questionnaire for that study condition. We then calculated the learning efficiency for each of the four conditions: (a) \textsc{fixed-amount} of guidance with \textsc{fixed-associations}, (b) \textsc{fixed-amount} of guidance with \textsc{adaptive-associations}, (c) \textsc{adaptive-amount} of guidance with \textsc{fixed-associations}, and (d) \textsc{adaptive-amount} of guidance with \textsc{adaptive-associations} using Formula~\ref{eq:2D}. This was done for both immediate recall (immediately after participants had completed the task) and delayed recall (a week after participants had completed the task). 

\subsubsection{Immediate Efficiency}
\par The immediate recall learning efficiency results across the study conditions are depicted in \autoref{fig:newtablegraphs3} (left). It shows that, for both \textsc{fixed-amount} and \textsc{adaptive-amount}, the \textsc{adaptive-associations} condition is more efficient (i.e., has a higher efficiency value) compared to the \textsc{fixed-associations} condition. 

\par The effects of the \textsc{amount} and \textsc{type} of guidance, on immediate recall learning efficiency are shown in \autoref{fig:newtablegraphs2} (top centre). 
The data is analysed using a mixed between-within subjects ANOVA on the 20\% trimmed means (\autoref{fig:newtablegraphs2} (bottom centre)).
There is no significant effect of the \textsc{amount} of guidance conditions on participants' immediate recall learning efficiency ($\mathit{df} = 16.80$, $p = 0.460$). 

\par By contrast, the effect of the \textsc{type} of instructions on immediate recall learning efficiency is statistically significant ($\mathit{df} = 16.80$, $p < 0.001$). \autoref{fig:newtablegraphs2} (top centre) indicates that the immediate recall learning efficiency in the \textsc{adaptive-associations} condition is significantly higher compared to the \textsc{fixed-associations} condition. The interaction effect of the \textsc{amount} and \textsc{type} on the immediate recall learning efficiency is not statistically significant ($\mathit{df} = 16.80$, $p > 0.05$).

\subsubsection{Delayed Efficiency}
\par The delayed recall learning efficiency results across the study conditions are shown in \autoref{fig:newtablegraphs3} (right). The figure indicates that, for both \textsc{fixed-amount} and \textsc{adaptive-amount}, the \textsc{adaptive-association} is more efficient (i.e., has a higher efficiency value) compared to the \textsc{fixed-associations} condition. 

\par The effects of the \textsc{amount} and \textsc{type} of guidance on delayed recall learning efficiency are illustrated in \autoref{fig:newtablegraphs2} (top right).
\par There is no significant effect of the \textsc{amount} of guidance condition on the delayed recall learning efficiency ($\mathit{df} = 16.72$, $p = 0.460$). 

\par By contrast, the effect of the \textsc{type} of instructions on the delayed recall learning efficiency is statistically significant ($\mathit{df} = 16.72$, $p < 0.001$). The results also show that  the delayed recall learning efficiency is significantly higher for \textsc{adaptive-associations} compared to \textsc{fixed-associations} condition. No statistically significant interaction effect was found between conditions ($\mathit{df} = 16.72$, $p > 0.05$).

\subsection{System Usability and User Experience}
The average SUS score shows that the system usability is in an acceptable range (\textsc{fixed-amount} = $82.3$, \textsc{adaptive-amount} = $85.8$; $SUS > 70$). The results for UEQ show that pragmatic (\textsc{fixed-amount} = $1.54$, \textsc{adaptive-amount} = $1.75$) and hedonic qualities (\textsc{fixed-amount} = $2.33$, \textsc{adaptive-amount} = $2.52$), as well as overall user experience (\textsc{fixed-amount} = $1.93$, \textsc{adaptive-amount} = $2.14$) are perceived as strongly positive in both conditions. In all cases we see that the adaptive-amount has higher values.

\subsection{Correlations between Measures}
\par A Pearson multiple correlation test was applied to find out whether or not there were any correlations between the measures (recall, efficiency, mental effort, task completion time). The results are presented in \autoref{tab:FPCTable}.

\begin{table*}[ht]
  \caption{Correlations between different measures for the \textsc{fixed-amount} of guidance condition (left) and the \textsc{adaptive-amount} of guidance condition (right). ME: mental effort, CT: task completion time, IR: immediate recall, DR: delayed recall, IE: immediate efficiency and DE: delayed efficiency.}
  \centering
  \begin{tabular}{c}
      \includegraphics[width=0.49\textwidth]{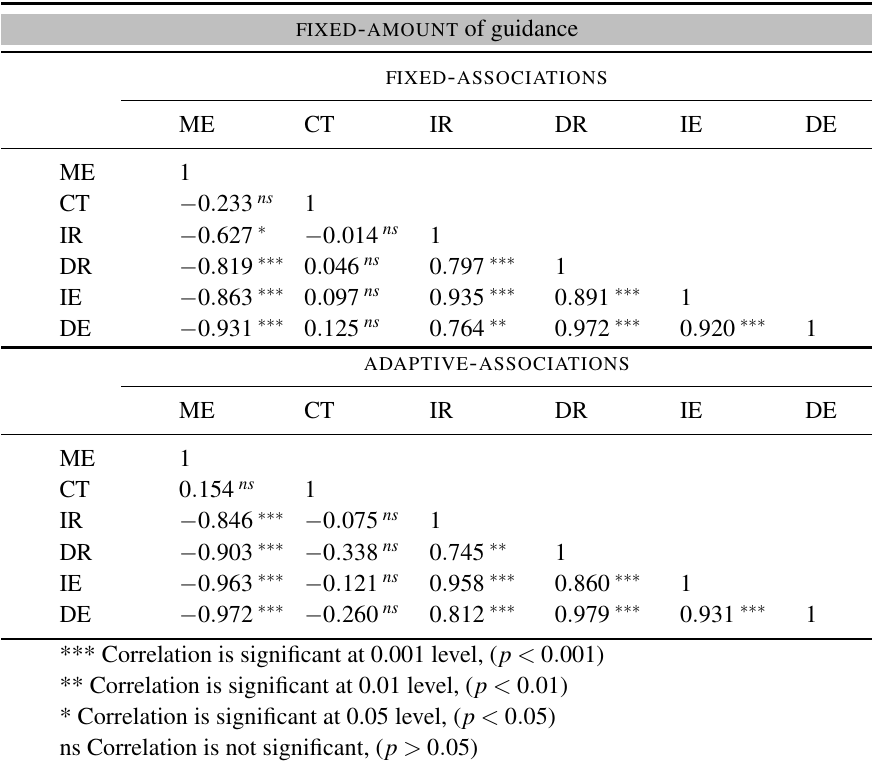} 
      \includegraphics[width=0.49\textwidth]{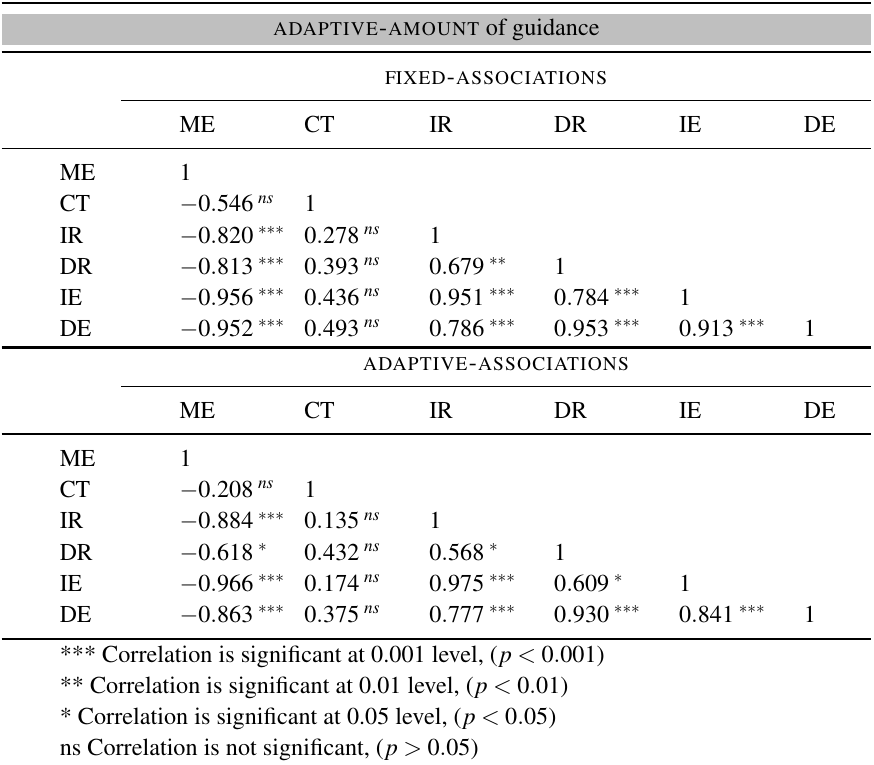} 
  \end{tabular}
  \label{tab:FPCTable}

\end{table*}

\par The results for the \textsc{fixed-amount} condition indicate that in both \textsc{fixed-associations} and \textsc{adaptive-associations} conditions, the mental effort is negatively related to both immediate and delayed recall as well as to both immediate and delayed efficiency. The task completion time measure does not show any relationship to other variables. The correlations between immediate/delayed and recall/efficiency variables was expected since these measures are all related.

\par Similar results can be observed for \textsc{adaptive-amount} condition. In both \textsc{fixed-associations} and \textsc{adaptive-associations} conditions, the mental effort is negatively related to both immediate and delayed recall as well as to both immediate and delayed efficiency. 
The task completion time does also not show any relationship to other variables. While, as before, immediate/delayed and recall/efficiency correlations are not surprising.


\section{Discussion}
\label{sec:discussion}
\par Arigat\={o} was developed to compare how learners perform in learning a new language, when presented with different adaptation factors of \rev{AR based} instructional guidance in an experiential learning environment. \rev{The AR system was deliberately selected as the most probable future technology that will be used in the classroom. An AR classroom has several advantages over other types of digitised classrooms such as taking the real world into consideration and consequently embedding information directly into the user’s field of view. These advantages better support real-world in-person communication and group collaboration compared to other technologies AR is often contrasted to, such as desktop, tablet computers, and VR~\cite{rzayev2020effectsar,morrison2009beesar}.
}

\rev{By taking the real world into consideration, our prototype did not just show content on the HMD, but allowed users to move and place AR objects in their physical surroundings. With this functionality, AR was used to replicate the real-world experiential learning scenario (e.g., learning by decorating a room for Christmas celebrations with AR objects as can be done with real objects) and show information in a coherent and meaningful way within the real world context.
}

Arigat\={o} allowed us to vary (i) the \textsc{amount} of guidance, which is either \textsc{fixed} or an \textsc{adaptive-amount} \rev{(the amount changes based on participants' performance)}, and (ii) the \textsc{type} of association, which is either \textsc{fixed} or with \textsc{adaptive-associations} \rev{(participants can select associations by themselves from a predefined set)}. \rev{We used a 2 × 2 mixed design to evaluate the study conditions, as explained in \autoref{sec:study_conditions}. The \textsc{amount} of guidance was evaluated as a within-subjects variable while the \textsc{type} of instructions as a between-subjects variable}. In this paper, the impact of the guidance method was measured through performance (i.e., recall and efficiency) and engagement (i.e., task completion time, mental effort, and motivation). 

\par We measured behavioural engagement for \textsc{fixed} guidance and \textsc{adaptive} guidance by observing how participants interacted with Arigat\={o} and how much time they spent in each condition. We could observe participants being more interested in the \textsc{adaptive-associations} than the \textsc{fixed-associations} condition, even though they completed the task in less time. This was expected, since keyword methods have been suggested as motivating learners to be more creative and to enjoy using their minds more productively~\cite{richardson1980}.

\par We measured cognitive engagement for all the study conditions by analysing how much mental effort participants invested in completing the task. We found that the mental effort with the \textsc{adaptive-amount} of guidance was significantly lower compared to the \textsc{fixed-amount} of guidance. This is in line with the literature, as prior work has shown that repeating the same information could increase the amount of cognitive work and consequently the mental effort~\cite{salomon1983}. We further identified that the mental effort in the \textsc{adaptive-associations} condition was lower than in \textsc{fixed-associations} condition. Again, this was somewhat expected, as the association method connects new learning to prior knowledge through the use of visual and/or acoustic cues~\cite{putnam2015,mastropieri1991}, which could lower the mental effort. However, we have shown that in the context of language learning the AR environment can support the creation of adaptive constructs not only in the form of the keywords (the ``keyword method''~\cite{atkinson1975}) and mental associations, but also in combination with 3D digital objects that users can manipulate with. It is this combination that supports experiential learning through manipulation of digital objects and that outperformed the traditional ``keyword method'' in our study. 

\par We measured learning performance through analysis of recall (immediate recall after the study and delayed recall test 7 days after the study) and learning efficiency. Interestingly, we found that compared to the \textsc{adaptive-amount} of guidance, the \textsc{fixed-amount} of guidance group scored better in the immediate and delayed recall tests, although they have invested a significantly higher mental effort during the task. In addition, we found that the \textsc{adaptive-associations} outperformed the \textsc{fixed-associations} group, in the immediate, delayed recall tests, and learning efficiency, while investing a significantly lower mental effort and spending a less amount of time during the task. These finding suggest avenues for future work in archetypal user models, which might allow users to slowly adapt pre-developed personal models from a standard set of ideal users. What we would refer to as the ``follow the leader'' learning approach. 

\par We have also investigated how different variables are correlated and how they could be combined in order to more accurately present engagement and learning performances of the participants' for each study condition. This contrasts to the existing literature in learning that often looks at these aspects individually. Our results show correlations between \textit{mental effort and learning performances (both recall and efficiency)} in all conditions, which was to be expected since the two are considered related. Interestingly, there were no correlations found between \textit{task completion time and learning performance} in any condition. Again, these findings suggest potential future work in ``transfer motivation'' between individual learners or between instructors and learners. 


\section{Limitations and future work}
\label{limitations}

\rev{The average age of participants in our study was 25 years, which presents a possible age bias. Despite this, this age group is worth studying as it is highly mobile, spending an extended period of time in a foreign speaking country (for example, the EU Erasmus+ programme alone funds more than half a million exchanges yearly~\cite{Erasmus}). As such, this group could benefit from an improved language learning system. Nevertheless, the results cannot be generalised over the whole population, and expanding the study to other age groups and exploring the effect of age on the proposed learning system is an important future direction.}

\rev{Another bias is the self-selection bias because the participants who opted-in are likely to be at ease when learning a new language. However, even if this study attracted only a particular group of language learners, the results still show the benefit of AR guidance with adaptive associations that would most likely benefit also other users in our studied age group. However, this needs to be further investigated. Using questionnaires for measuring certain aspects such as motivation and mental effort can result in social desirability bias as users could answer them to please the researcher. However, we used well established questionnaires together with measurements that can hardly be affected by such bias (i.e., recall and efficiency and task completion time).}

\rev{The Arigat\={o} prototype covers the first three levels (categories) of the Bloom's taxonomy~\cite{bloom2001revised} only: Knowledge (remembering the words, sentences), Comprehension (understanding the structure of sentences), and Application (applying or speaking out the sentences).  The expansion to higher levels of the taxonomy (Analysis (analyse), Synthesis, and Evaluation (evaluate), Creation) is also relevant and we aim to take this into consideration in future implementations. In addition, we used the task of learning vocabulary and other language constructs as a use-case to explore the potential of AR in adaptive learning scenarios. In future,  this use-case should be expanded to other use-cases to confirm its generalisability. }


\section{Conclusion}

In this paper we present the Arigat\={o} (Augmented Reality Instructional Guidance \& Tailored Omniverse) prototype -- an adaptive guidance augmented reality (AR) system for language learning. AR has been identified as an ideal platform to support simulating various experiences for experiential learning \rev{and we explored the AR's design space in the learning context rather than comparing it to other technologies}. With our prototype we investigated how the amount of guidance (fixed vs.\ adaptive-amount) and the type of guidance (fixed vs.\ adaptive-associations) affects the engagement and consequently the learning outcomes of language learning in an AR environment. 

Compared to the adaptive-amount, the fixed-amount of guidance group scored better in the immediate and delayed (after 7 days) recall tests. However, this group also invested a significantly higher mental effort to complete the task. 
Adaptive-associations group outperformed the fixed-associations group in the immediate, delayed (after 7 days) recall tests, and learning efficiency. The adaptive-associations group also showed significantly lower mental effort and spent less time to complete the task. 
Both results hint at potential for the archetypal user models in comparable learning scenarios, which might allow users to ``transfer motivation'' between individual learners or between instructors and learners, and slowly adapt pre-developed personal models. Such approach could achieve the balance between the (adaptive) amount and (adaptive) type to optimise the mental effort need to complete the learning task.

The results also show the potential of AR in adaptive learning scenarios where the learning environment needs to be simulated and where virtual content needs to be added or removed on the fly based on the learning needs. Learning vocabulary and other language constructs as a use-case explored in this study, presents just one of many possibilities that offer venues for future work \rev{in AR classrooms as it is likely that the technology will get better and widely accessible in the future~\cite{liono2021}.}

\balance

\bibliographystyle{abbrv-doi}

\bibliography{arXiv}

\begin{thebibliography}{100}

\bibitem{bloom2001revised}
A taxonomy for learning, teaching, and assessing; a revision of bloom's
  taxonomy of educational objectives, complete edition, 2001.

\bibitem{Yaz2009}
{\em Implicit emotional tagging of multimedia using EEG signals and brain
  computer interface}, 2009.

\bibitem{herbert2017}
{\em An Adaptive AR Tutor For Cabling a Network Topology.}, 2017.

\bibitem{alrashidi2016}
O.~Alrashidi, H.~P. Phan, and B.~H. Ngu.
\newblock Academic engagement: An overview of its definitions, dimensions, and
  major conceptualisations.
\newblock {\em International Education Studies}, 9(12):41--52, 2016.

\bibitem{adams2008}
American Institute of Physics.
\newblock {\em What levels of guidance promote engaged exploration with
  interactive simulations?}, vol. 1064, 2008.

\bibitem{amiryousefi2011}
M.~Amiryousefi and S.~Ketabi.
\newblock Mnemonic instruction: A way to boost vocabulary learning and recall.
\newblock {\em Journal of Language Teaching \& Research}, 2(1), 2011.

\bibitem{anderson1985}
J.~R. Anderson, C.~F. Boyle, and B.~J. Reiser.
\newblock Intelligent tutoring systems.
\newblock {\em Science}, 228(4698):456--462, 1985.

\bibitem{App2008}
J.~J. Appleton, S.~L. Christenson, and M.~J. Furlong.
\newblock Student engagement with school: Critical conceptual and
  methodological issues of the construct.
\newblock {\em Psychology in the Schools}, 45(5):369--386, 2008.

\bibitem{arvanitis2021}
P.~Arvanitis and P.~Krystalli.
\newblock Digital storytelling and augmented reality-based scenarios for
  foreign language teaching.
\newblock In {\em Digital Pedagogies and the Transformation of Language
  Education}, pp. 1--22. IGI Global, 2021.

\bibitem{ary2018}
D.~Ary, L.~C. Jacobs, C.~K.~S. Irvine, and D.~Walker.
\newblock {\em Introduction to research in education}.
\newblock Cengage Learning, 2018.

\bibitem{atkinson1975}
R.~C. Atkinson.
\newblock Mnemotechnics in second-language learning.
\newblock {\em American psychologist}, 30(8):821, 1975.

\bibitem{birt2018}
J.~Birt, Z.~Stromberga, M.~Cowling, and C.~Moro.
\newblock Mobile mixed reality for experiential learning and simulation in
  medical and health sciences education.
\newblock {\em Information}, 9(2):31, 2018.

\bibitem{boud1993}
D.~Boud, R.~Cohen, and D.~Walker.
\newblock {\em Using experience for learning}.
\newblock McGraw-Hill Education (UK), 1993.

\bibitem{brooke1996}
J.~Brooke et~al.
\newblock Sus-a quick and dirty usability scale.
\newblock {\em Usability evaluation in industry}, 189(194):4--7, 1996.

\bibitem{bruner1961}
J.~S. Bruner.
\newblock The act of discovery.
\newblock {\em Harvard educational review}, 1961.

\bibitem{canhoto2016}
A.~I. Canhoto and J.~Murphy.
\newblock Learning from simulation design to develop better experiential
  learning initiatives: An integrative approach.
\newblock {\em Journal of Marketing Education}, 38(2):98--106, 2016.

\bibitem{carbonneau2015}
K.~J. Carbonneau and S.~C. Marley.
\newblock Instructional guidance and realism of manipulatives influence
  preschool children's mathematics learning.
\newblock {\em The Journal of Experimental Education}, 83(4):495--513, 2015.

\bibitem{clark2011}
R.~C. Clark, F.~Nguyen, and J.~Sweller.
\newblock {\em Efficiency in learning: Evidence-based guidelines to manage
  cognitive load}.
\newblock John Wiley \& Sons, 2011.

\bibitem{clark2009}
R.~E. Clark.
\newblock How much and what type of guidance is optimal for learning from
  instruction.
\newblock {\em Constructivist instruction: Success or failure}, pp. 158--183,
  2009.

\bibitem{cohen1980}
A.~D. Cohen and E.~Aphek.
\newblock Retention of second-language vocabulary overtime: Investigating the
  role of mnemonic associations.
\newblock {\em System}, 8(3):221--235, 1980.

\bibitem{cohen1988}
J.~Cohen.
\newblock Statistical power analysis for the behavioral sciences.
\newblock {\em England: Routledge}, 1988.

\bibitem{cronbach1977}
L.~J. Cronbach and R.~E. Snow.
\newblock {\em Aptitudes and instructional methods: A handbook for research on
  interactions.}
\newblock Irvington, 1977.

\bibitem{dita2016}
F.-A. Dita.
\newblock A foreign language learning application using mobile augmented
  reality.
\newblock {\em Informatica Economica}, 20(4):76, 2016.

\bibitem{draxler2020}
F.~Draxler, A.~Labrie, A.~Schmidt, and L.~L. Chuang.
\newblock Augmented reality to enable users in learning case grammar from their
  real-world interactions.
\newblock In {\em Proceedings of the 2020 CHI Conference on Human Factors in
  Computing Systems}, pp. 1--12, 2020.

\bibitem{dunlosky2013}
J.~Dunlosky, K.~A. Rawson, E.~J. Marsh, M.~J. Nathan, and D.~T. Willingham.
\newblock Improving students’ learning with effective learning techniques:
  Promising directions from cognitive and educational psychology.
\newblock {\em Psychological Science in the Public interest}, 14(1):4--58,
  2013.

\bibitem{ellis199}
N.~C. Ellis.
\newblock The psychology of foreign language vocabulary acquisition:
  Implications for call.
\newblock {\em Computer Assisted Language Learning}, 8(2-3):103--128, 1995.

\bibitem{Erasmus}
{Erasmus+}.
\newblock Factsheets and statistics on {Erasmus}+.
\newblock \url{https://erasmus-plus.ec.europa.eu/node/2585}.
\newblock 2022-05-24.

\bibitem{ferrero2018}
G.~Ferrero, F.~Bichai, and M.~Rusca.
\newblock Experiential learning through role-playing: Enhancing stakeholder
  collaboration in water safety plans.
\newblock {\em Water}, 10(2):227, 2018.

\bibitem{Fra2010}
C.~A. Frantzidis, C.~Bratsas, C.~L. Papadelis, E.~Konstantinidis, C.~Pappas,
  and P.~D. Bamidis.
\newblock Toward emotion aware computing: an integrated approach using
  multichannel neurophysiological recordings and affective visual stimuli.
\newblock {\em IEEE transactions on Information Technology in Biomedicine},
  14(3):589--597, 2010.

\bibitem{fredricks2004}
J.~A. Fredricks, P.~C. Blumenfeld, and A.~H. Paris.
\newblock School engagement: Potential of the concept, state of the evidence.
\newblock {\em Review of educational research}, 74(1):59--109, 2004.

\bibitem{freund2011}
P.~A. Freund, J.-T. Kuhn, and H.~Holling.
\newblock Measuring current achievement motivation with the qcm: Short form
  development and investigation of measurement invariance.
\newblock {\em Personality and Individual Differences}, 51(5):629--634, 2011.

\bibitem{grantcharov2004}
T.~P. Grantcharov, V.~B. Kristiansen, J.~Bendix, L.~Bardram, J.~Rosenberg, and
  P.~Funch-Jensen.
\newblock Randomized clinical trial of virtual reality simulation for
  laparoscopic skills training.
\newblock {\em British journal of surgery}, 91(2):146--150, 2004.

\bibitem{guan2021}
J.-Q. Guan, L.-H. Wang, Q.~Chen, K.~Jin, and G.-J. Hwang.
\newblock Effects of a virtual reality-based pottery making approach on junior
  high school students’ creativity and learning engagement.
\newblock {\em Interactive Learning Environments}, pp. 1--17, 2021.

\bibitem{halabi2006}
A.~K. Halabi.
\newblock Applying an instructional learning efficiency model to determine the
  most efficient feedback for teaching introductory accounting.
\newblock {\em Global Perspectives on Accounting Education}, 3(1):6, 2006.

\bibitem{herman2009}
P.~Herman and L.~M. Gomez.
\newblock Taking guided learning theory to school: Reconciling the cognitive,
  motivational, and social contexts of instruction.
\newblock pp. 74--93, 2009.

\bibitem{huang2021}
G.~Huang, X.~Qian, T.~Wang, F.~Patel, M.~Sreeram, Y.~Cao, K.~Ramani, and A.~J.
  Quinn.
\newblock Adaptutar: An adaptive tutoring system for machine tasks in augmented
  reality.
\newblock In {\em Proceedings of the 2021 CHI Conference on Human Factors in
  Computing Systems}, pp. 1--15, 2021.

\bibitem{huang2016}
T.-C. Huang, C.-C. Chen, and Y.-W. Chou.
\newblock Animating eco-education: To see, feel, and discover in an augmented
  reality-based experiential learning environment.
\newblock {\em Computers \& Education}, 96:72--82, 2016.

\bibitem{ibrahim2018}
A.~Ibrahim, B.~Huynh, J.~Downey, T.~H{\"o}llerer, D.~Chun, and J.~O'donovan.
\newblock Arbis pictus: A study of vocabulary learning with augmented reality.
\newblock {\em IEEE transactions on visualization and computer graphics},
  24(11):2867--2874, 2018.

\bibitem{Tak2003}
IEEE.
\newblock {\em Remarks on emotion recognition from multi-modal bio-potential
  signals}, vol.~2, 2003.

\bibitem{jantjies2018}
M.~Jantjies, T.~Moodley, and R.~Maart.
\newblock Experiential learning through virtual and augmented reality in higher
  education.
\newblock In {\em Proceedings of the 2018 international conference on education
  technology management}, pp. 42--45, 2018.

\bibitem{jarmon2009}
L.~Jarmon, T.~Traphagan, M.~Mayrath, and A.~Trivedi.
\newblock Virtual world teaching, experiential learning, and assessment: An
  interdisciplinary communication course in second life.
\newblock {\em Computers \& Education}, 53(1):169--182, 2009.

\bibitem{joplin1981}
L.~Joplin.
\newblock On defining experiential education.
\newblock {\em Journal of experiential education}, 4(1):17--20, 1981.

\bibitem{joseph2013measuring}
S.~Joseph.
\newblock {\em Measuring cognitive load: A comparison of self-report and
  physiological methods}.
\newblock PhD thesis, Arizona State University, 2013.

\bibitem{kalyuga2001l}
S.~Kalyuga, P.~Chandler, and J.~Sweller.
\newblock Learner experience and efficiency of instructional guidance.
\newblock {\em Educational Psychology}, 21(1):5--23, 2001.

\bibitem{Kat2008}
C.~D. Katsis, N.~Katertsidis, G.~Ganiatsas, and D.~I. Fotiadis.
\newblock Toward emotion recognition in car-racing drivers: A biosignal
  processing approach.
\newblock {\em IEEE Transactions on Systems, Man, and Cybernetics-Part A:
  Systems and Humans}, 38(3):502--512, 2008.

\bibitem{Kea1998}
G.~Kearsley and B.~Shneiderman.
\newblock Engagement theory: A framework for technology-based teaching and
  learning.
\newblock {\em Educational technology}, 38(5):20--23, 1998.

\bibitem{kirschner2006}
P.~A. Kirschner, J.~Sweller, and R.~E. Clark.
\newblock Why minimal guidance during instruction does not work: An analysis of
  the failure of constructivist, discovery, problem-based, experiential, and
  inquiry-based teaching.
\newblock {\em Educational psychologist}, 41(2):75--86, 2006.

\bibitem{koedinger2007}
K.~R. Koedinger and V.~Aleven.
\newblock Exploring the assistance dilemma in experiments with cognitive
  tutors.
\newblock {\em Educational Psychology Review}, 19(3):239--264, 2007.

\bibitem{Koe2012}
R.~A.~L. Koelstra.
\newblock Affective and implicit tagging using facial expressions and
  electroencephalography., 2012.

\bibitem{kolb1984}
D.~Kolb.
\newblock {\em Experiential learning: experience as the source of learning and
  development}.
\newblock Prentice Hall, 1984.

\bibitem{kolb1981}
D.~A. Kolb.
\newblock Experiential learning theory and the learning style inventory: A
  reply to freedman and stumpf.
\newblock {\em Academy of Management Review}, 6(2):289--296, 1981.

\bibitem{kolb2001}
D.~A. Kolb, R.~E. Boyatzis, C.~Mainemelis, et~al.
\newblock Experiential learning theory: Previous research and new directions.
\newblock {\em Perspectives on thinking, learning, and cognitive styles},
  1(8):227--247, 2001.

\bibitem{kuder1937}
G.~F. Kuder and M.~W. Richardson.
\newblock The theory of the estimation of test reliability.
\newblock {\em Psychometrika}, 2(3):151--160, 1937.

\bibitem{kuhn2007}
D.~Kuhn.
\newblock Is direct instruction an answer to the right question?
\newblock {\em Educational psychologist}, 42(2):109--113, 2007.

\bibitem{lazonder2016}
A.~W. Lazonder and R.~Harmsen.
\newblock Meta-analysis of inquiry-based learning: Effects of guidance.
\newblock {\em Review of educational research}, 86(3):681--718, 2016.

\bibitem{lewis2009}
J.~R. Lewis and J.~Sauro.
\newblock The factor structure of the system usability scale.
\newblock In {\em International conference on human centered design}, pp.
  94--103. Springer, 2009.

\bibitem{lewis1994}
L.~H. Lewis and C.~J. Williams.
\newblock Experiential learning: Past and present.
\newblock {\em New directions for adult and continuing education},
  1994(62):5--16, 1994.

\bibitem{liono2021}
R.~A. Liono, N.~Amanda, A.~Pratiwi, and A.~A. Gunawan.
\newblock A systematic literature review: learning with visual by the help of
  augmented reality helps students learn better.
\newblock {\em Procedia Computer Science}, 179:144--152, 2021.

\bibitem{liu2013}
P.-H.~E. Liu and M.-K. Tsai.
\newblock Using augmented-reality-based mobile learning material in efl english
  composition: An exploratory case study.
\newblock {\em British journal of educational technology}, 44(1):E1--E4, 2013.

\bibitem{lu2015}
S.-J. Lu and Y.-C. Liu.
\newblock Integrating augmented reality technology to enhance children’s
  learning in marine education.
\newblock {\em Environmental Education Research}, 21(4):525--541, 2015.

\bibitem{mair2020}
P.~Mair and R.~Wilcox.
\newblock Robust statistical methods in r using the wrs2 package.
\newblock {\em Behavior research methods}, 52(2):464--488, 2020.

\bibitem{mastropieri1991}
M.~A. Mastropieri and T.~E. Scruggs.
\newblock {\em Teaching students ways to remember: Strategies for learning
  mnemonically.}
\newblock Brookline Books, 1991.

\bibitem{mather2017}
C.~Mather, T.~Barnett, V.~Broucek, A.~Saunders, D.~Grattidge, and W.~Huang.
\newblock Helping hands: using augmented reality to provide remote guidance to
  health professionals.
\newblock In {\em Context Sensitive Health Informatics: Redesigning Healthcare
  Work}, pp. 57--62. IOS Press, 2017.

\bibitem{mayer2009}
R.~E. Mayer.
\newblock Constructivism as a theory of learning versus constructivism as a
  prescription for instruction.
\newblock {\em Constructivist instruction: Success or failure}, pp. 184--200,
  2009.

\bibitem{mergel1998}
B.~Mergel.
\newblock Instructional design and learning theory, 1998.

\bibitem{merrill2002}
M.~D. Merrill.
\newblock First principles of instruction.
\newblock {\em Educational technology research and development}, 50(3):43--59,
  2002.

\bibitem{moorhouse2017}
N.~Moorhouse, T.~Jung, et~al.
\newblock Augmented reality to enhance the learning experience in cultural
  heritage tourism: An experiential learning cycle perspective.
\newblock {\em eReview of Tourism Research}, 8, 2017.

\bibitem{moorhouse2019}
N.~Moorhouse, M.~C. tom Dieck, and T.~Jung.
\newblock An experiential view to children learning in museums with augmented
  reality.
\newblock {\em Museum Management and Curatorship}, 34(4):402--418, 2019.

\bibitem{morrison2009beesar}
A.~Morrison, A.~Oulasvirta, P.~Peltonen, S.~Lemmela, G.~Jacucci, G.~Reitmayr,
  J.~N\"{a}s\"{a}nen, and A.~Juustila.
\newblock Like bees around the hive: A comparative study of a mobile augmented
  reality map.
\newblock In {\em Proceedings of the SIGCHI Conference on Human Factors in
  Computing Systems}, CHI '09, p. 1889–1898. Association for Computing
  Machinery, New York, NY, USA, 2009. doi: {{%
10\hspace{.1pt}\discretionary{.}{%
}{.}\hspace{.4pt}1145\discretionary{/}{%
}{/}1518701\hspace{.1pt}\discretionary{.}{%
}{.}\hspace{.4pt}1518991}}


\bibitem{mueller2003}
D.~Mueller and J.~M. Ferreira.
\newblock Marvel: A mixed-reality learning environment for vocational training
  in mechatronics.
\newblock In {\em Proceedings of the Technology Enhanced Learning International
  Conference (TEL 03)}, 2003.

\bibitem{paas1993}
F.~G. Paas and J.~J. Van~Merri{\"e}nboer.
\newblock The efficiency of instructional conditions: An approach to combine
  mental effort and performance measures.
\newblock {\em Human factors}, 35(4):737--743, 1993.

\bibitem{paivio1981}
A.~Paivio and A.~Desrochers.
\newblock Mnemonic techniques in second-language learning.
\newblock {\em Journal of Educational Psychology}, 73(6):780, 1981.

\bibitem{pla1983}
R.~L. Plackett.
\newblock Karl pearson and the chi-squared test.
\newblock {\em International Statistical Review/Revue Internationale de
  Statistique}, pp. 59--72, 1983.

\bibitem{pressley1982}
M.~Pressley, J.~R. Levin, and H.~D. Delaney.
\newblock The mnemonic keyword method.
\newblock {\em Review of Educational Research}, 52(1):61--91, 1982.

\bibitem{putnam2015}
A.~L. Putnam.
\newblock Mnemonics in education: Current research and applications.
\newblock {\em Translational Issues in Psychological Science}, 1(2):130, 2015.

\bibitem{raugh1975}
M.~R. Raugh and R.~C. Atkinson.
\newblock A mnemonic method for learning a second-language vocabulary.
\newblock {\em Journal of Educational Psychology}, 67(1):1, 1975.

\bibitem{reeve2011}
J.~Reeve and C.-M. Tseng.
\newblock Agency as a fourth aspect of students’ engagement during learning
  activities.
\newblock {\em Contemporary Educational Psychology}, 36(4):257--267, 2011.

\bibitem{rheinberg2001}
F.~Rheinberg, R.~Vollmeyer, and B.~D. Burns.
\newblock Fam: Ein fragebogen zur erfassung aktueller motivation in lern-und
  leistungssituationen (langversion, 2001).
\newblock {\em Diagnostica}, 2:57--66, 2001.

\bibitem{richardson1980}
J.~T. Richardson and A.~Jones.
\newblock {\em Mental imagery and human memory}.
\newblock Macmillan International Higher Education, 1980.

\bibitem{rzayev2020effectsar}
R.~Rzayev, S.~Korbely, M.~Maul, A.~Schark, V.~Schwind, and N.~Henze.
\newblock Effects of position and alignment of notifications on ar glasses
  during social interaction.
\newblock In {\em Proceedings of the 11th Nordic Conference on Human-Computer
  Interaction: Shaping Experiences, Shaping Society}. Association for Computing
  Machinery, New York, NY, USA, 2020.

\bibitem{Pie2017}
SAGE Publications Sage CA: Los Angeles, CA.
\newblock {\em Engagement and Competence in VR and non-VR Environments},
  vol.~61, 2017.

\bibitem{salomon1983}
G.~Salomon.
\newblock The differential investment of mental effort in learning from
  different sources.
\newblock {\em Educational psychologist}, 18(1):42--50, 1983.

\bibitem{Sch2013}
W.~B. Schaufeli.
\newblock What is engagement.
\newblock {\em Employee engagement in theory and practice}, 15:321, 2013.

\bibitem{Ueq2019}
M.~Schrepp.
\newblock Ueq-user experience questionnaire, 2019.

\bibitem{schrepp2017}
M.~Schrepp, A.~Hinderks, and J.~Thomaschewski.
\newblock Design and evaluation of a short version of the user experience
  questionnaire (ueq-s).
\newblock {\em IJIMAI}, 4(6):103--108, 2017.

\bibitem{schuman1981}
H.~Schuman, S.~Presser, and J.~Ludwig.
\newblock Context effects on survey responses to questions about abortion.
\newblock {\em Public Opinion Quarterly}, 45(2):216--223, 1981.

\bibitem{seedhouse2014}
P.~Seedhouse, A.~Preston, P.~Olivier, J.~Dan, H.~Philip, B.~Madeline,
  A.~Rafiev, and M.~Kipling.
\newblock The european digital kitchen project.
\newblock {\em Bellaterra Journal of Teaching \& Learning Language \&
  Literature}, 7(1):1--16, 2014.

\bibitem{sha1965}
S.~S. Shapiro and M.~B. Wilk.
\newblock An analysis of variance test for normality (complete samples).
\newblock {\em Biometrika}, 52(3/4):591--611, 1965.

\bibitem{Mur2008}
Springer.
\newblock {\em Time-frequency analysis of EEG signals for human emotion
  detection}, 2008.

\bibitem{steffe1995}
L.~P. Steffe and J.~E. Gale.
\newblock {\em Constructivism in education}.
\newblock Psychology Press, 1995.

\bibitem{sullivan2013}
G.~M. Sullivan and A.~R. Artino~Jr.
\newblock Analyzing and interpreting data from likert-type scales.
\newblock {\em Journal of graduate medical education}, 5(4):541--542, 2013.

\bibitem{int2016}
L.~Sullivan.
\newblock Interquartile range (iqr), 2016.

\bibitem{thoravi2019}
B.~Thoravi~Kumaravel, F.~Anderson, G.~Fitzmaurice, B.~Hartmann, and
  T.~Grossman.
\newblock Loki: Facilitating remote instruction of physical tasks using
  bi-directional mixed-reality telepresence.
\newblock In {\em Proceedings of the 32nd Annual ACM Symposium on User
  Interface Software and Technology}, pp. 161--174, 2019.

\bibitem{tobias1982}
S.~Tobias.
\newblock When do instructional methods.
\newblock {\em Educational Researcher}, 11(4):4--9, 1982.

\bibitem{tobias2009}
S.~Tobias and T.~M. Duffy.
\newblock {\em Constructivist instruction: Success or failure?}
\newblock Routledge, 2009.

\bibitem{Top2019}
F.~B. Topu and Y.~Goktas.
\newblock The effects of guided-unguided learning in 3d virtual environment on
  students' engagement and achievement.
\newblock {\em Computers in Human Behavior}, 92:1--10, 2019.

\bibitem{vaughan2017}
K.~L. Vaughan, R.~E. Vaughan, and J.~M. Seeley.
\newblock Experiential learning in soil science: Use of an augmented reality
  sandbox.
\newblock {\em Natural Sciences Education}, 46(1):1--5, 2017.

\bibitem{vazquez2017}
C.~D. Vazquez, A.~A. Nyati, A.~Luh, M.~Fu, T.~Aikawa, and P.~Maes.
\newblock Serendipitous language learning in mixed reality.
\newblock In {\em Proceedings of the 2017 CHI Conference Extended Abstracts on
  Human Factors in Computing Systems}, pp. 2172--2179, 2017.

\bibitem{Web2019}
M.~Weber, J.~Giacomin, A.~Malizia, L.~Skrypchuk, V.~Gkatzidou, and
  A.~Mouzakitis.
\newblock Investigation of the dependency of the drivers’ emotional
  experience on different road types and driving conditions.
\newblock {\em Transportation research part F: traffic psychology and
  behaviour}, 65:107--120, 2019.

\bibitem{weerasinghe2019}
M.~Weerasinghe, A.~Quigley, J.~Ducasse, K.~{\v{C}}. Pucihar, and M.~Kljun.
\newblock Educational augmented reality games.
\newblock pp. 3--32, 2019.

\bibitem{wei2016}
X.~Wei, D.~Weng, Y.~Liu, and Y.~Wang.
\newblock A tour guiding system of historical relics based on augmented
  reality.
\newblock In {\em 2016 IEEE Virtual Reality (VR)}, pp. 307--308. IEEE, 2016.

\bibitem{westerfield2015}
G.~Westerfield, A.~Mitrovic, and M.~Billinghurst.
\newblock Intelligent augmented reality training for motherboard assembly.
\newblock {\em International Journal of Artificial Intelligence in Education},
  25(1):157--172, 2015.

\bibitem{weston1986}
C.~Weston and P.~A. Cranton.
\newblock Selecting instructional strategies.
\newblock {\em The Journal of Higher Education}, 57(3):259--288, 1986.

\bibitem{wickens2015}
C.~D. Wickens, J.~G. Hollands, S.~Banbury, and R.~Parasuraman.
\newblock {\em Engineering psychology and human performance}.
\newblock Psychology Press, 2015.

\bibitem{wise2009}
A.~F. Wise and K.~O'Neill.
\newblock Beyond more versus less: A reframing of the debate on instructional
  guidance.
\newblock 2009.

\bibitem{worthen2011}
J.~B. Worthen and R.~R. Hunt.
\newblock {\em Mnemonology: Mnemonics for the 21st century}.
\newblock Psychology Press, 2011.

\bibitem{yang2018}
S.~Yang and B.~Mei.
\newblock Understanding learners’ use of augmented reality in language
  learning: insights from a case study.
\newblock {\em Journal of Education for Teaching}, 44(4):511--513, 2018.

\end{thebibliography}
\end{document}